\documentclass[sigconf]{acmart}

\AtBeginDocument{%
  \providecommand\BibTeX{{%
    \normalfont B\kern-0.5em{\scshape i\kern-0.25em b}\kern-0.8em\TeX}}}

\copyrightyear{2022}
\acmYear{2022}
\setcopyright{rightsretained}
\acmConference[CHI '22]{CHI Conference on Human Factors in Computing Systems}{April 29-May 5, 2022}{New Orleans, LA, USA}
\acmBooktitle{CHI Conference on Human Factors in Computing Systems (CHI '22), April 29-May 5, 2022, New Orleans, LA, USA}
\acmDOI{10.1145/3491102.3517470}
\acmISBN{978-1-4503-9157-3/22/04}

\usepackage{color}
\usepackage{textcomp}
\usepackage{multirow}
\usepackage{subcaption}
\usepackage{bm}
\usepackage{wrapfig}
\newcommand{\xhdr}[1]{\vspace{1mm}\noindent{{\bf #1.}}}

\newcommand{\new}[1]{\textcolor{black}{#1}}
\newcommand{\newn}[1]{\textcolor{black}{#1}}
\begin{document}

%%
%% The "title" command has an optional parameter,
%% allowing the author to define a "short title" to be used in page headers.
\title[From Who You Know to What You Read]{From Who You Know to What You Read: \newn{Augmenting Scientific Recommendations with Implicit Social Networks}}

\author{Hyeonsu B. Kang}
\orcid{0000-0002-1990-2050}
\affiliation{%
  \institution{Carnegie Mellon University}
  \streetaddress{5000 Forbes Ave}
  \city{Pittsburgh}
  \state{PA}
  \country{USA}
}
\email{hyeonsuk@cs.cmu.edu}

\author{Rafal Kocielnik}
\orcid{0000-0001-5602-6056}
\affiliation{%
  \institution{University of Washington}
  \streetaddress{1400 NE Campus Parkway}
  \city{Seattle}
  \state{WA}
  \postcode{98195}
  \country{USA}
}
\email{rkoc@uw.edu}
\author{Andrew Head}
\orcid{0000-0002-1523-3347}
\affiliation{%
  \institution{Allen Institute for AI}
  \streetaddress{2157 N Northlake Way \#110}
  \city{Seattle}
  \state{WA}
  \postcode{98103}
  \country{USA}
}
\email{andrewhead@allenai.org}
\author{Jiangjiang Yang}
\affiliation{%
  \institution{Allen Institute for AI}
  \streetaddress{2157 N Northlake Way \#110}
  \city{Seattle}
  \state{WA}
  \postcode{98103}
  \country{USA}
}
\email{jjyang@allenai.org}
\author{Matt Latzke}
\affiliation{%
  \institution{Allen Institute for AI}
  \streetaddress{2157 N Northlake Way \#110}
  \city{Seattle}
  \state{WA}
  \postcode{98103}
  \country{USA}
}
\email{mattl@allenai.org}
\author{Aniket Kittur}
\orcid{0000-0003-4192-9302}
\affiliation{%
  \institution{Carnegie Mellon University}
  \streetaddress{5000 Forbes Ave}
  \city{Pittsburgh} 
  \state{PA} 
  \postcode{15213}
  \country{USA}
}
\email{nkittur@cs.cmu.edu}

\author{Daniel S. Weld}
\orcid{0000-0002-3255-0109}
\affiliation{%
  \institution{Allen Institute for AI \& \\University of Washington}
   \streetaddress{2157 N Northlake Way \#110}
  \city{Seattle}
  \state{WA}
  \postcode{98103}
  \country{USA}
}
\email{danw@allenai.org}
\author{Doug Downey}
\orcid{0000-0002-4737-8444}
\affiliation{%
  \institution{Allen Institute for AI}
  \streetaddress{2157 N Northlake Way \#110}
  \city{Seattle}
  \state{WA}
  \postcode{98103}
  \country{USA}
}
\affiliation{%
  \institution{Northwestern University}
  \streetaddress{1400 NE Campus Parkway}
  \city{Evanston}
  \state{IL}
  \country{USA}
}
\email{dougd@allenai.org}
\author{Jonathan Bragg}
\affiliation{%
  \institution{Allen Institute for AI}
  \streetaddress{2157 N Northlake Way \#110}
  \city{Seattle}
  \state{WA}
  \postcode{98103}
  \country{USA}
}
\email{jbragg@allenai.org}

\renewcommand{\shortauthors}{H. B. Kang, R. Kocielnik, A. Head, J. Yang, M. Latzke, A. Kittur, D. Weld, D. Downey, and J. Bragg}

%%
%% The abstract is a short summary of the work to be presented in the
%% article.
\begin{abstract}

The ever-increasing pace of scientific publication necessitates methods for quickly identifying relevant papers. While neural recommenders trained on user interests can help, they still result in long, monotonous lists of suggested papers. To improve the discovery experience we introduce multiple new methods for \new{\em augmenting} recommendations with textual relevance messages that highlight knowledge-graph connections between recommended papers and a user's publication and interaction history. We explore associations mediated by author entities and those using citations alone. In a large-scale, real-world study, we show how our approach significantly increases engagement---and future engagement when mediated by authors---without introducing bias towards highly-cited authors. To expand message coverage for users with less publication or interaction history, we develop a novel method that highlights connections with proxy authors of interest to users and evaluate it in a controlled lab study. Finally, we synthesize design implications for future graph-based messages.

\end{abstract}

%%
%% Teaser figure
% \begin{teaserfigure}
%   \includegraphics[width=\linewidth]{figures/teaser_placeholder.png}
%   \caption{Teaser figure placeholder}
%   ~\label{fig:system}
%   \vspace{-2em}
% \end{teaserfigure}

%%
%% This command processes the author and affiliation and title
%% information and builds the first part of the formatted document.
\maketitle

%%
%% The code below is generated by the tool at http://dl.acm.org/ccs.cfm.
%% Please copy and paste the code instead of the example below.
%%


%% Keywords. The author(s) should pick words that accurately describe
%% the work being presented. Separate the keywords with commas.

\section{Introduction}
Keeping on top of the literature while not being overwhelmed is a fundamental yet aspirational goal for many scientists today~\cite{pain2016keep,woolston2019phds,williams2019postgraduate,hemp2009death} due to the immense scale of archival knowledge and its continued growth ~\cite{jinha2010article,bornmann2015growth,noorden_doubling_output,tenopir2009electronic}. While paper recommendation systems can help users identify useful papers from the larger literature, users can have difficulty understanding why those papers might be relevant to them or match their interests. \new{Existing a}pproach\new{es} for addressing this challenge \new{include providing} explanations of the recommender's behavior\new{, or providing relevance messages that supplement recommendations, which may increase the persuasiveness and informativeness of the recommendations}~\cite{tintarev2007survey,wang2018explainable,zhao2019personalized,zhao2014we,zhang2018explainable,smith2020no,symeonidis2008providing,sato2019action,he2015trirank,sharma2013social,bilgic2005explaining,herlocker2000explaining,tintarev2008effectiveness,toderici2010finding,wang2018ripplenet}.

%However, it is unclear how these techniques may be adapted to support scientific recommendations and whether they would lead to meaningful differences in scientists' engagement. 

%Scientific recommender systems can help but they too result in long, monotonous lists of papers that induce disengagement from the users. The fundamental challenge is convincing users why a recommended paper is worth their attention to click on.

In this paper we compare the effectiveness of various types of relevance messages for scientific paper recommendations in a large-scale randomized study. Previous work has suggested the potential effectiveness of citation~\cite{bethard2010should}-, knowledge~\cite{catherine2016personalized,chaudhari2017entity,ostuni2013top}-, and social-~\cite{sharma2013social} graphs for recommendations and \new{relevance messages} in various \new{domains of recommendations}. However, what type of \new{messages} is most impactful for scientific paper recommendations remains an open question. \newn{To answer this,} we compare two types of \new{messages} that expose information about the interaction between the citation graph and the user. The first type of \new{messages} finds the intersection between reference papers in new paper recommendations---which are recently published and often do not have any incoming citations or rich metadata available---and papers the user previously published or personally curated \new{(e.g., `This paper cites 2 papers in your library'). We refer to this approach as citation-based relevance messaging that leverages} \textit{relevance via what they read}. Our second kind infers the implicit social network of authors from the citation graph to feature author-focused connections to recommended papers. \new{Specifically, this inferred author network consists of relations targeted to our domain, such as which authors previously co-authored together or who has cited whose work} \new{(e.g., `John Doe authored 3 papers you cited'). We refer to this approach as direct author-based relevance messaging that leverages} \textit{relevance via who they know}. Unlike prior work that used citation~\cite{bethard2010should}-, knowledge~\cite{catherine2016personalized,chaudhari2017entity,ostuni2013top}-, and social-graphs~\cite{sharma2013social} to recommend or explain the recommendations, we use \new{the citation and inferred author networks} strictly to generate useful relevance messages \new{that supplement the recommendations, in order to} support a broad set of information needs that goes beyond the maximal content relevance~\cite{case2016looking,belkin1985interaction}. \new{Our relevance messages are generated} in a model-agnostic and post-hoc manner, \newn{which may generalize to new application contexts} \new{independent of the underlying recommender algorithms}. To study users' authentic engagement patterns in a real-world scenario, we conducted \new{a} \newn{field} experiment \new{on} an existing alert system of a popular scholarly search engine, which sends out emails with personalized paper recommendations to users who have opted in to alerts.

Through an iterative design process, we developed robust message designs for use in our two-month-long study with over {7,000} participants. Comparing the emails featuring direct author-based relevance messages to emails featuring citation-based relevance messages and Control, we saw \newn{the largest} significant increases in user engagement from the emails augmented with \new{direct author-based} messages, which had \new{user click-through rates} \newn{that were 28\%} higher than Control. \new{Direct a}uthor-based relevance messages also seemed to result in a higher level of future engagement, with 13\% overall increase over Control in the email open rates, \new{and 30\% increase after the first two-week exposure to the messages.} Furthermore, through an analysis of the distribution of clicked paper recommendations, we found that user engagement did not shift to papers written by authors with higher academic status when \new{direct author-based messages} were used compared to Control, suggesting that the messages were unlikely to exacerbate the \textit{rich-get-richer} phenomenon~\cite{merton1968matthew}. 

However, the effectiveness of \new{direct} author-based relevance messages was limited by their scarcity; they boosted engagement more than citation-based relevance messages despite occurring much less frequently, on 4\% of paper recommendations compared to 9\%.
%While only less than 4\% of paper recommendations featured author-based relevance messages, the rate was significantly higher for citation-based messages, at around 9\%.
In follow-up analyses using generalized linear mixed models, we show that substantially increasing the \% of paper recommendations featured with \new{direct} author-based relevance messages is indeed a mechanism likely effective for further engaging users, even after controlling for potential \new{covariates}. To \new{increase the coverage}, we designed and implemented a new method for expanding the relevance relations on the implicit social network using {\em indirect} author-based relevance messages, which borrow from the networks of potentially familiar and trusted middle authors \new{via} [author]-[trusted author]-[user] triplet relations (e.g., \new{assuming Dr. Anthony Fauci is a user-trusted middle author,} ``Catherine Paules has authored 4 papers that Dr. Anthony Fauci cited.; You saved 5 of Dr. Anthony Fauci's papers in the library.''). In other words, the indirect author-based relevance incorporates \newn{implicit} `endorsement' from a known intermediate author, \new{Dr. Fauci, whom} the user may trust and from whom they may appreciate paper recommendations.

In a controlled lab study with fourteen scientists, we show the feasibility of indirect author-based messages for increasing message coverage.
%by featuring familiar indirect authors in the message that users can recognize to make sense of the relevance conveyed in the message.
In addition, we gained qualitative insights into different types of benefits and challenges involved with \new{different types of} relevance messages. At a high-level, the benefits can be classified into two categories. The first category is benefits directly \textit{on} recommendations, such as mobilizing the user's mental models of authors to gain a deeper understanding of recommended papers, or judging the potential usefulness of them. The second category of benefits is rather \textit{around} the recommendations, such as developing awareness of other scientists they care about, understanding connections within academic communities, and understanding one's impact in the academic community. These findings confirm results from prior studies that cast recommendation as a socially embedded process that depends on both trust and the relationship of individuals~\cite{lueg1997social,perugini2004recommender}, but also surface new factors and design implications specific to scientific recommendations situated in a broader intellectual community. 

In summary, this work makes five contributions. First, we designed and implemented two types of graph-based \new{(citation and inferred author network)} relevance messages for augmenting personalized paper recommendations, grounded in an iterative design process and interviews with multiple stakeholders. Second, we present evidence from a large-scale online deployment study showing that our messaging approaches indeed increased user engagement, and that \new{direct} author-based relevance messages performed the best, although their full potential was likely not reached due to low coverage. Third, we designed and implemented an additional, indirect author-based message that models endorsement from intermediate authors who are likely known and trusted, in order to further engage users by \new{expanding the relations found on the inferred author network} to mitigate the scarcity of author-based relevance messages. Fourth, through a controlled lab study with fourteen scientists, we show the feasibility of indirect author-based relevance messages and present qualitative insights into how different types of relevance messages benefited users. Finally, we present design implications for future augmentation approaches that aim to incorporate graph-based relevance information.

\section{Related Work}
\subsection{\new{Exploratory Search Needs in Scholarly Recommendations}} \label{subsection:current_practices}
The information environment for today's scientists can be characterized by the immense scale and dynamic changes~\cite{jinha2010article,bornmann2015growth,noorden_doubling_output,tenopir2009electronic,rowland2002overcoming,megwalu2015academic,hole2008email,whittaker1996email}, which make effective allocation of attention~\cite{simon1996designing} an imperative for scientists. The issue of information overload~\cite{miller1960information} is especially pronounced in the crucial task of staying up-to-date with the relevant literature~\cite{landhuis2016scientific}. Compounded with domain-specific knowledge barriers that make the scholarly reading experience challenging~\cite{hillesund2010digital,bazerman1985physicists,nicholas2010researchers}, scientists' experience with the literature is often characterized as tedious, scattered, and relying upon chance discovery~\cite{breitinger2019too}. 

\new{Exploratory search and curatorial needs for scholarly recommendations are high in this environment~\cite{landhuis2016scientific}. Additionally, what scientists judge as relevant may not only be logical---such as topical or narrowly defined as papers containing specific terms---but also situational, and depend on the scientist's personal information needs that go beyond the need for maximal content relatedness~\cite{case2016looking,belkin1985interaction}. %\newn{However, this notion of relevance is underserved by scholarly search engines or is too costly to.
To support the exploratory process of curating high-relevance papers, scientists \newn{commonly} adopt two kinds of \newn{often effortful} strategies. The first kind can be characterized by its use of \textit{citation} networks which enable scientists to search forward or backward in time through citation or reference chaining~\cite{barrett2005information,covi1999material,vakkari2006searching,westbrook2003information,bates2002speculations,jia2017analysis}. \newn{Chaining in this manner requires scientists to maintain the relevance and coherence between papers as their number grows exponentially.} The second kind leverages a different type of network, which is \textit{social} in nature, \newn{to support serendipitous} sharing and finding of recommendations on social media platforms such as  Twitter~\cite{mohammadi2018academic,holmberg2014disciplinary,vainio2017highly,fang2021science,klar2020using} or cold-emailing high-profile experts in an outside field to receive valuable bibliography~\cite{vakkari2006searching,covi1999material,palmer2005scholarly,palmer2009scholarly}.}

\new{To support these needs while reducing the burden of the laborious process, \newn{prior work} has explored ways to automatically explain the recommended items' relevance to the users. Such explanations of relevance may be generated after the recommendations are derived~\cite{zhang2018explainable,sato2019action}, and may be focused on providing explanations of the recommendation mechanisms themselves, which can better engage users by enhancing their understanding of the inner workings of the `black-box' recommender algorithms~\cite{smith2020no,symeonidis2008providing}. In contrast, \newn{rather than focusing on providing faithful explanations of the inner algorithm, our work seeks to} increase the persuasiveness and informativeness~\cite{bilgic2005explaining,sharma2013social} of recommendations \newn{by incorporating \textit{external} relevance signals}. Existing work in this space has explored ways to incorporate relevant knowledge graph entities~\cite{catherine2016personalized,chaudhari2017entity,ostuni2013top} or social signals such as local (e.g., showing the user's immediate friends' preference) or global (e.g., overall popularity) relevance~\cite{sharma2013social} to increase persuasiveness. However, open questions remain as to how and what kinds of relevance information may be incorporated into scholarly recommendations to effect large-scale behavioral changes in the scientist user population, and which approach works best.}

\vspace{-1em}
\subsection{Behavior Change and Motivation}
% High level arguments for novelty:
% 1. explored in health and few others, but not in scholarly + impact of task and population in incentives
% 2. most applied in artificial AMT settings, lab studies, or limited deployments in artificial tasks + translation of theory and lab into real-world is not trivial
% 3. Most studied of short-term effectiveness, not long-term, which limits the generalizability + novelty wears off
% 4. Personalized persuasion based on real-world data, rather than personality traits evaluated with surveys.
\new{To study the large-scale changes among the scientists engaging with scholarly recommendations, our work builds upon the literature on behavior change and persuasion. These areas suggests several techniques, such as principles of authority and social influence~\cite{cialdini1987influence, ajzen1991theory, cialdini2004social}, which inspired our designs for increasing user engagement.
While these theories have been adapted in practice, our work is different from all prior explorations in key aspects.}

\new{The vast majority of prior work tested their motivation strategies only with crowd-sourced populations on platforms such as Amazon Mechanical Turk \cite{grau2018personalized, kocielnik2017send, mcinnis2016one, hsieh2016you}.
Furthermore, most message-based behavior change designs have been applied in the context health \& well-being \cite{muench2017more, kocielnik2017send, jang2020healthier}, civic engagement \cite{grau2018personalized}, UI testing \cite{mcinnis2016one}, or creative, brainstorming tasks \cite{hsieh2016you}. Demographics and incentives of users on crowd-sourcing platforms and in personal contexts can be substantially different from our professional scholarly population \cite{traunmueller2015crowdsourcing}.
Finally, translating theory-informed or only lab-tested motivation designs into real-world systems has been shown to be a non-trivial task due to feasibility limitations and ethical considerations~\cite{grau2018personalized, colusso2019translational}. Aside from these general limitations of prior work, we further discuss in detail the specific differences in relation to selected works closest to our designs.}

% another limitation: personalization in prior work relies on personality traits derived via surveys - not feasible in real-world, limited automation

\new{Grau et al.~\cite{grau2018personalized} designed  motivation-supportive messages in the context of a crowd-civic platform, to engage volunteers, yet these designs did not explore leveraging \emph{relevance} or \emph{social graph}, focusing instead on aspects of \emph{controlled} and \emph{autonomous} motivation tested only with a limited sample of the Korean population. Kocielnik and Hsieh~\cite{kocielnik2017send} designed message triggers that are diversified either by cognitively close concepts to the targeted action or the recipient, and found that the close-to-recipient diversification was more effective. This work offers insights about the importance of \emph{closeness}, yet it was tested in a limited deployment with 27 participants and in a substantially different domain of physical activity. Unfortunately, factors affecting engagement can be substantially different across different settings and populations \cite{durham2005health, ben2018investigating}. McInnis et al.~\cite{mcinnis2016one} focused on motivating one-time comments in an artificial task of ``testing website interface'' and hence is substantially different from motivating long-term engagement. Hsieh et al.~\cite{hsieh2016you} focused on exploring differences in user populations attracted by different incentives (e.g., monetary reward vs lottery), but did not explore the design of messaging or leveraging social graphs and was limited to motivating single-time study participation.}
%It also focuses on motivating one-time study participation rather than long-term engagement in live scholarly system.

\vspace{-1.5em}
\subsection{\new{Social Network-based Relevance Information}} \label{subsection:social_relevance}
\new{Social recommendation has been a millenia-old mechanism for finding personally relevant items, but the advent of social media enabled its propagation at much greater scale. Social recommendations are also common in the scholarly domain as described above; scientists use platforms such as Twitter to share and access relevant papers to read. Theoretically, important factors of social recommendation include homophily~\cite{mcpherson2001birds}, diffusion and influence~\cite{singla2008yes,leskovec2006patterns}, and trust~\cite{ma2009learning,liu2010incorporating}, which originate from social network analysis. Prior work in social network-centric recommendations developed designs targeted at such factors to improve user engagement~\cite{guy2009personalized,sharma2011network}. However, open questions remain as to how the (social) network information may be incorporated in scholarly recommendations. No explicit network of scientists exists, and the interesting network structure may not even be inherently social in the sense studied in prior research. For example, while several works have studied the notion of \textit{immediate} friends in network-centric recommendations and relevance explanations based on the theory of homophily (e.g., close friend groups may share similar tastes in music) and trust (e.g., `if my friend likes this album, it must be good') (e.g.,~\cite{sharma2013social,sharma2011network,guy2009personalized}), the importance of immediate relationships may fade away rather quickly in the scholarly domain. The insight here is that similar kinds of immediate closeness may not be as useful or even interesting to scientist users, as they may already be familiar with the work by their `friends' (e.g., through co-authorship) and may instead benefit more from relevant yet novel work mediated through \newn{more distant} connections. This implies characteristics of relevance diffusion and trust relations within scholarly networks may differ from those of the domains previously studied. Furthermore, important questions arise from the perspective of system-wide fairness, for example whether frequently featuring distant yet trusted authors results in negative externalities such as skewed distribution of work visibility and subsequent downstream reduction of impact from the work made relatively less visible (e.g., the Matthew effect~\cite{merton1968matthew}).}

\new{In this work, we contribute to the literature of social network-centric relevance messages in the domain of scholarly recommendations. We expand the design space by introducing \newn{a novel messaging technique that leverages an intermediate, trusted author to expand the coverage of relevance connections beyond the user's own history, in order to highlight additional papers or authors that may be of interest to the user}. We contribute a large-scale deployment study and a controlled lab study evaluating the effectiveness of messages on inducing behavior and motivation changes. We further uncover the distinctive usefulness and challenges of social network-centric relevance messages. We end with a discussion of open questions remaining for future work.}

\section{Relevance Message Design for Scientific Recommendations}
Following an iterative design process,
we designed two strategies for \new{engaging} users \new{with email alerts containing paper recommendations}: \emph{citation-} and \emph{author-based relevance messages}, \new{which augment the recommendations}.
In this section, we describe our scientific recommendation setting and message design iterations, and present a detailed description of the designs themselves.

\subsection{Email Alerts for Scientific Paper Recommendations}
\label{sec:sys}
\begin{figure}
    \centering
    \includegraphics[width=.45\textwidth]{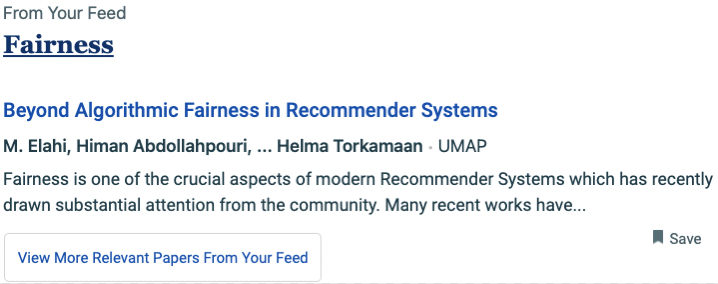}
    \vspace{-1em}
    \caption{An abbreviated email alert.}
    \vspace{-1.5em}
    \label{fig:email_alert}
\end{figure}

We designed our engagement strategies to work within email alerts sent from Semantic Scholar. The Semantic Scholar platform provided users the ability to search and curate interesting papers for personalized research feeds. \new{Users could explicitly assign a binary positive (`more like this') or negative (`less like this') rating to each paper in their curated research feeds.} Users then could opt in to receive alert emails when new relevant papers are found for each feed on a regular basis, with a user-selected frequency (e.g., daily or weekly). Typical alert emails started with the title of the feed, and a list of new relevant papers (usually ranging from 1 to 50; Fig.~\ref{fig:email_alert} shows a simplified version). Emails included basic paper metadata and link-based affordances for navigating to and saving papers, managing alert subscriptions, etc. Working within an existing production paper alert system lent ecological validity to our experiments. Due to the email setting, our design space did not include complex interactions such as hovers. The research feeds recommend papers using a neural recommender, trained on the user's individual paper ratings. Importantly, our relevance messages are agnostic to the underlying recommendation model, \new{which was held constant across the conditions.}
%This setting provided us an ideal ground for testing how users might react to different types of relevance messages with high task realism, because paper recommendations were personalized to individual users as predicted by a neural recommender to contain genuinely interesting and relevant content for personally curated feeds. Therefore, it provided us an experimental condition with high ecological validity to observe users' authentic engagement patterns.\jb{Shorten and describe how explanations are recommendation model-agnostic}
%\subsubsection{Initial Motivations} 
%- persuasion, behavior change - author importance, penalization
%- recommendation systems - relevance
%- social graph in scholar network - social network inspirations
\subsection {Iterative Design Process}
Our designs went through four phases of iterations and prototyping. 
%In phase 1 we brainstormed a theory informed set of engagement strategies. In phase 2 we narrowed down these strategies based on availability of data and technical feasibility. In phase 3 we refined the content and visual design following feedback from marketing and design teams. Finally, in phase 4, we ran an internal beta-testing on a live system with $\sim100$ internal users and conducted interviews with 7 to address technical issues and refine the design further. %Here we describe the details of these phases.

% Initial brainstorming: https://docs.google.com/document/d/1fPLMEbdjAwtTH3_TJcyI4km1IOWgSLJKqPDIwJUe-F8/edit?usp=sharing
\textit{Phase 1 - Theory \& Expert driven brainstorming.} Design brainstorming took place among three authors (two with knowledge of scholarly recommendation services and one with expertise in behavior change and persuasion) informed by literature from behavior change \& persuasion (e.g., principles of authority and social influence) \cite{cialdini1987influence, ajzen1991theory, cialdini2004social}, recommendation systems (e.g., relevance and discovery) \cite{anderson2020algorithmic}, and information processing literature (summarizing, balancing and diversifying the information) \cite{johnson2015information, pirolli1999information}. Several strategies were eliminated on the grounds of: 1) challenging execution in deployment context, 2) data availability, and 3) ethics (e.g., scarcity, which may present a false impression of limited availability).

\textit{Phase 2 - GUI prototyping, data availability \& technical feasibility evaluation.} Selected strategies were prototyped in high visual and data model fidelities \cite{dhillon2011visual}. The visual fidelity prototyping explored text-based and visualization-based designs \cite{kocielnik2017send, fernandes2018uncertainty}. The high data model fidelity prototyping ensured: 1) the designs worked as intended in the live system, 2) the complete record of interactions was preserved, and 3) a sufficient number of users were impacted on a daily basis. 

\textit{Phase 3 - Graphical Design, Marketing \& Engineering Feedback.} Additional graphic design review ensured the use of an icon, font, and color scheme was consistent with the existing alert emails. Further marketing team feedback resulted in rephrasing parts of the messages to fit factual \& information-centric tone (e.g., an early phrase \textit{``Cites:''} was replaced with \textit{``Also cites:''} to more factually reflect the presented numbers). Engineering team feedback led to improved data pooling, freshness and completeness.

% Beta testing user feedback - https://docs.google.com/document/d/1wzmGXPojhjolcaGLcZAkcfoNkdXVUINQoqrhapI4OtI/edit?usp=sharing

\textit{Phase 4 - Beta testing.} The implemented designs were beta-tested for two weeks with 100 internal users in actual live service to: 1) eliminate any technical issues and 2) collect feedback from users in more naturalistic setting. Interviews were carried out with seven users over Slack and via a video call, which confirmed the general understandability, visibility, and user interest in the presented relevance messages. Several changes were introduced following user feedback: 1) dropping a change in email title, as it was not noticed, 2) replacing the separator between message parts from `+' sign to comma, and 3) clarifying or removing \newn{unclear} content. 

\vspace{-1em}
\subsection{Explanation Designs}
We present the detailed description of the final designs.
%Source for figures: https://docs.google.com/presentation/d/1GHMJyUrsNAOni4HazsLp9uNoq4WXHiT6EbUNKChyZoY/edit?usp=sharing
%Long horizontal figures used in the paper - in separate presentation because of different slide length: https://docs.google.com/presentation/d/1HE4o_DbRDvGgsR4GwjHzyCddD3jzMqVVVBE4F86Cbj4/edit?usp=sharing

\begin{figure*}[t]
    \includegraphics[width=.95\textwidth]{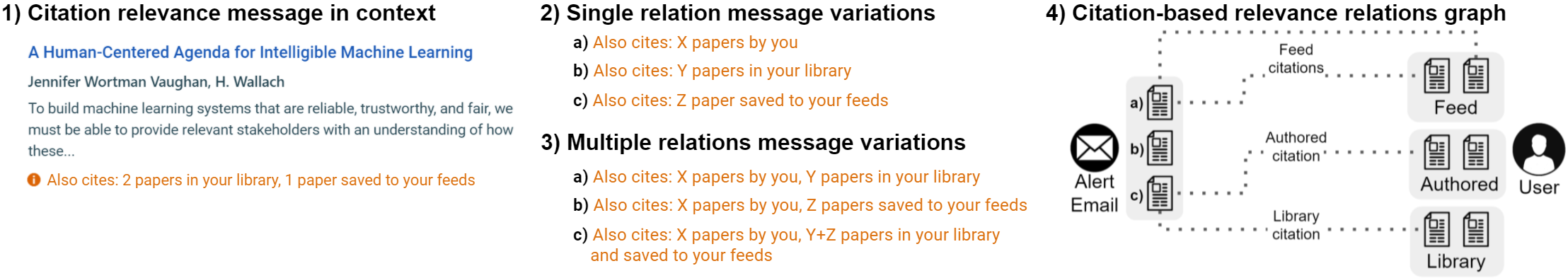}
    \vspace{-1.5em}
    \caption{\textbf{Citation relevance message design.} 1) An example of the relevance message rendered in paper context as shown to the user. 2) Examples of messages featuring relevance relations to different user papers: papers a) user authored, b) placed in the user's library, c) added to the user's feed. 3) Examples of messages featuring multiple relations. 4) Graph based depiction of the citation-based relevance messages: an alert paper a) citing two papers from the user's feed, b) with no relevance relation - it will have no message, c) citing both a user authored paper and a paper the user added to their library.}
    \label{fig:condition1_combined}
    \vspace{-1em}
\end{figure*}

\xhdr{Design 1: Citation-based Relevance Messages} {{\textit{Goal}}}: Design 1 conveys potential relevance of an alert paper to the user via direct citations (Fig \ref{fig:condition1_combined}.4). It aims to highlight citations \newn{from the alert paper to} papers the user has previously explicitly expressed interest in. \newn{We do not consider citations of the alert paper, since most alert papers are new and do not yet have many citations.} Prior work in scholarly contexts has emphasized the important of citations as a simple measure of relevance~\cite{aksnes2009researchers, hassan2017identifying}.

\textit{Design}: We define \emph{user relevant sources} to be the user's personal \emph{library} and \emph{research feed}, as well as papers the user has \emph{authored}. These sources contain papers \newn{in which} the user has explicitly expressed interest at some point. We further define a \emph{relevant alert paper} to be a paper that cites one or more of the papers in the user relevant sources. Our \emph{citation-based relevance condition} adds one of the relevance messages presented in Figs \ref{fig:condition1_combined}.2 and \ref{fig:condition1_combined}.3 to any relevant alert paper. If an alert paper does not cite any of the papers in the user relevant sources, no relevance message is added. Alert papers that cite papers from only one user relevant source receive one of the message variations presented in Fig.~\ref{fig:condition1_combined}.2, while alert papers citing papers from multiple user relevant sources receive message variations presented in Fig.~\ref{fig:condition1_combined}.3. Each message communicates information about the number of cited papers from a particular user relevant source, as well as provenance information about the source (e.g., library or feed) being cited. In a particular case when a paper cites papers from all the user relevant sources, a shorter variant of the message, which combines the citation counts for library and feed papers, is presented in order to reduce user cognitive load~\cite{sweller2011cognitive} (see Fig.~\ref{fig:condition1_combined}.3c).
% The \textit{``Also cites''} phrase reflects the fact that some papers included in the alert emails are already presented due to citing authors or papers user explicitly followed via these alert emails.

\begin{figure*}[t]
    \includegraphics[width=.95\textwidth]{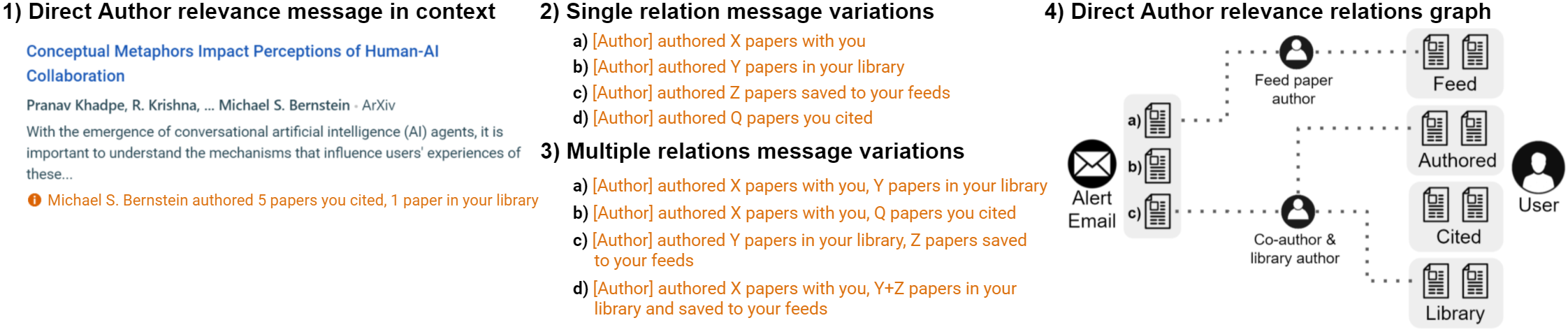}
    \vspace{-1.5em}
    \caption{\textbf{Direct Author relevance message design.} 1) An example of the message as shown in context to the user. 2) Examples of messages featuring author relevance relations to different user papers: papers user a) authored, b) added to the library, c) added to their feed, and d) cited. 3) Examples of messages featuring multiple relations. 4) Graph based depiction of the direct author-based messages:  an alert paper a) featuring an author of one paper from a user feed, b) with no direct author relevance relation (thus, having no message), and c) featuring an author of both a user-authored paper and a paper in the user's library.
    %\jb{Why co-author and library grouped? Thought we combined feed+library as interaction data.} \rafal{That's correct we only combined counts for feed+library, what made that impression?}
    }
    \label{fig:condition2_combined}
    \vspace{-1em}
\end{figure*}

\xhdr{Design 2: Direct Author-based Relevance Messages} \textit{Goal}: Design 2 conveys potential relevance of an alert paper to the user using their implicit social network of authors~(Fig.~\ref{fig:condition2_combined}.4). In particular the authors with papers that the user has explicitly expressed interest in are emphasized. The strategy is motivated by several indications from prior work that academics actively search for researchers with similar interests~\cite{megwalu2015academic}, but existing services offer limited support for such discovery \cite{breitinger2019too}. Further motivation for emphasizing author information comes from indications about the importance of author networks in academia~\cite{radford2020people} and the value of social networks of trusted sources~\cite{chen2010short,andersen2008trust} in broader recommendation contexts.

\textit{Design}: We define a relevant alert paper to be a paper authored by at least one \emph{user relevant author}, defined as having authored one or more papers in the user relevant sources. For this design, we expanded the number of user relevant sources to four by adding papers cited by the user. Early user feedback indicated that this source was meaningful only in Design 2. Our \emph{direct author-based relevance condition} adds one of the relevance messages presented in Figs.~\ref{fig:condition2_combined}.2 and \ref{fig:condition2_combined}.3 to any relevant alert paper. If none of the authors on an alert paper are user relevant authors, no message is added. If a paper was authored by more than one user relevant author, only the author with the highest number of papers in the user relevant sources is shown, with ties broken randomly. Alert papers featuring a user relevant author who has authored papers in one or more user relevant sources are featured with one of the message variations presented in Fig.~\ref{fig:condition2_combined}.2 or Fig.~\ref{fig:condition2_combined}.3, respectively. Each relevance message communicates information about the number of papers the user relevant author has authored in each user relevant source. If a featured author has authored papers in all the user relevant sources, the library and feeds sources are merged to create a shorter variant of the message and reduce user cognitive load \cite{sweller2011cognitive} (see Fig.~\ref{fig:condition2_combined}.3d).

\section{Study 1 - Large-scale online deployment study}
\subsection{Procedure} We randomly assigned over seven thousand email-alert users (see Section~\ref{sec:sys} for a system overview) to one of the three conditions: Control/No message (status quo experience with the research feed and new paper alerts), Citation (citation-based relevance messages are added), and Direct Author (author-based relevance messages are added). \new{Over a span of about two-months (April 7th -- June 13th, 2021), participants received regular alert emails containing new paper recommendations for personally curated research feeds.} Table~\ref{table:email_stats} shows descriptive statistics \new{of the dataset collected for analysis}. The system logged user engagement (e.g., opening an email or clicking a paper title to view the paper detail page on the search engine). \new{Thus, our measure of user engagement is two-fold, with the click-through rates grouped at the email level as a measure of engagement---i.e., \textsc{ctr}: a binary measure of either 1: \newn{the email was opened and at least one included} paper recommendation was clicked, or 0: otherwise \newn{(includes unopened emails)}---and an additional measure of \textit{future} engagement via email open rates. If participants found the emails useful, they would \newn{likely} open more of them in the future, resulting in higher overall email open rates. In this sense, higher email open rates are indicative of increased future engagement.}
\begin{table}
    \centering
    \begin{tabular}{c c c}
    \toprule
    \multirow{2}{*}{\textbf{Condition}} & \multirow{2}{*}{\textbf{\# of Users}} & \textbf{\newn{\# of Emails out of total}} \\
    & & \newn{featuring at least 1 message (\%)} \\
    \midrule
    Control & {2,248} & \newn{\textsc{n/a} out of {22,548}}\\
    Citation & {2,474} & \newn{{5,984} out of {23,658} (25\%)}\\
    Direct Author & {2,316} & \newn{{3,895} out of {22,657} (16\%)}\\
    \bottomrule
    \end{tabular}
    \caption{Statistics of the users and emails in our analysis.}
    \label{table:email_stats}
    \vspace{-3em}
\end{table}
\subsection{Results}
We report results from our studies below. \new{We compare the different messaging \textit{strategies} in our analyses rather than messages alone, i.e., the effect of messages together with the recommendations they augment, by comparing the full email dataset (unless otherwise specified). In such analyses, we use \textit{messages} interchangeably with \textit{messaging strategies}}. To denote statistical significance we use the following notations: $\alpha=.05 (^{*})$, $.01 (^{**})$, $.001 (^{***})$, $.0001(^{****})$. Alpha levels were adjusted when appropriate in post-hoc analyses using Bonferroni correction.
 
\subsubsection{\new{Both types of messaging strategies increased \textsc{ctr}, but only direct author-based messaging increased future engagement}}
\begin{figure*}[t]
    \begin{minipage}{0.45\linewidth}
        \begin{subfigure}[t]{0.45\linewidth}
            \centering
            \includegraphics[height=3cm]{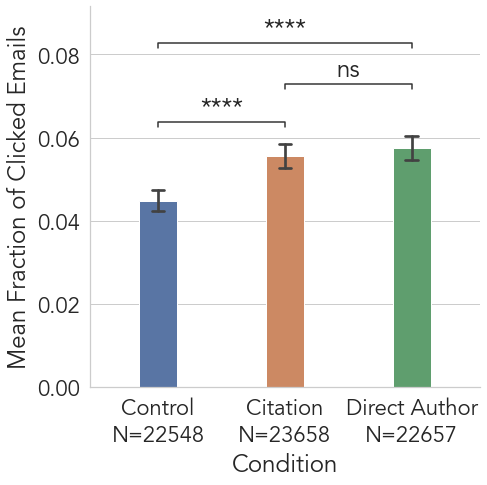}
            \caption{Both direct author ($\mu=0.058$) and citation ($\mu=0.056$) messaging resulted in significantly higher \textsc{ctr} than Control ($\mu=0.045$).}
            \vspace{-1em}
            \label{fig:click_rates}
        \end{subfigure}
        \quad
        \begin{subfigure}[t]{0.45\linewidth}
            \centering
            \includegraphics[height=3cm]{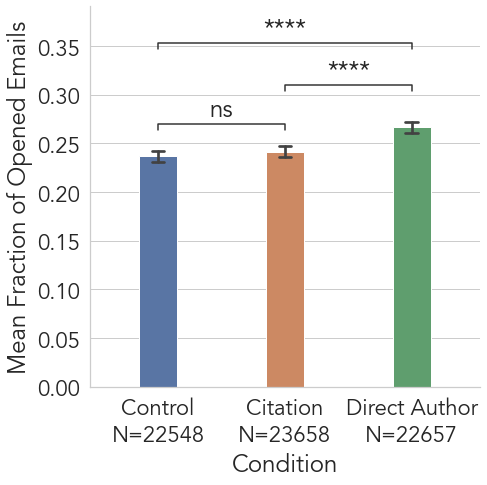}
            \caption{Direct author messaging ($\mu=0.27$) resulted in the highest email open rates  ($\mu=0.24$ for both Citation and Control).}
            \vspace{-1em}
            \label{fig:open_rates}
        \end{subfigure}
        \caption{Analysis of the \textsc{ctr} and email open rates by type.}
    \end{minipage}
    \qquad
    \begin{minipage}{0.45\linewidth}
        \begin{subfigure}[t]{0.45\linewidth}
            \centering
            \includegraphics[height=3cm]{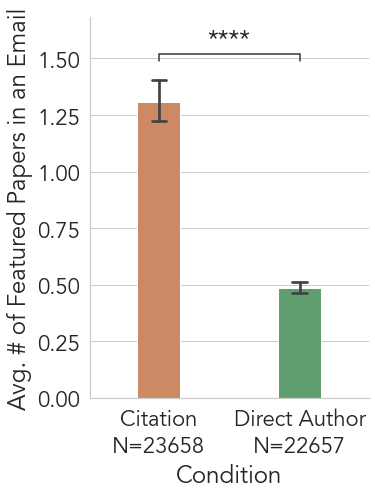}
            \caption{Direct author messages were significantly rarer ($\mu=0.5$) than Citation messages ($\mu=1.3$) in each email.}
            \vspace{-1em}
            \label{fig:avg_num_featured_papers}
        \end{subfigure}
        \quad
        \begin{subfigure}[t]{0.45\linewidth}
            \centering
            \includegraphics[height=3cm]{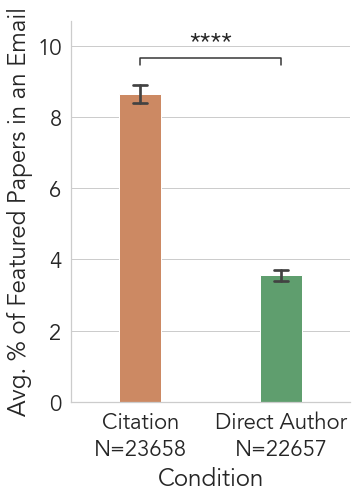}
            \caption{The rate of messages showed a similar pattern ($\mu=8$\% for Citation vs. $\mu=4$\% for Direct Author messages).}
            \vspace{-1em}
            \label{fig:avg_frac_featured_papers}
        \end{subfigure}
        \caption{Analysis of the message prevalence by type.}
    \end{minipage}
    \vspace{-1em}
\end{figure*}
We found that both types of relevance messages significantly increased the \textsc{ctr} over Control, with Direct Author ($\mu=0.058, \bar{\sigma}=0.2330$, $t(44622.22)=6.20, p=5.6\times10^{-10}$) and Citation ($\mu=0.056, \bar{\sigma}=0.2292$, $t(46065.34)=5.36, p=8.3\times10^{-8}$) resulting in higher \textsc{ctr}s than Control ($\mu=0.045, \bar{\sigma}=0.2068$) (Welch's two-tailed t-test, Fig.~\ref{fig:click_rates}). However, the difference was not significant between the Citation and Direct Author conditions ($t(46149.07)=0.92, p=0.36$). \new{We validated these results with analyses on potential biases of randomization, such as the average email length (Fig.~\ref{fig:mean_email_length}) and message position (Fig.~\ref{fig:mean_of_mean_message_position}) (see Appendix A)}.

We further examined whether different message types resulted in differences in how often users open the emails. The Direct Author condition showed a significantly higher email open rate than the rest ($\mu=0.27, \bar{\sigma}=0.442$ vs. $\mu=0.24, \bar{\sigma}=0.428$ in Citation and $\mu=0.24, \bar{\sigma}=0.425$ in Control, Fig.~\ref{fig:open_rates}, significant at $\alpha=1.0\times 10^{-4}$). Because the subjects, headers, and other metadata of emails did not differ between conditions, it is likely that the difference in open rates is attributable to the content of the emails and specifically the type of relevance messages featured in them.

\new{Furthermore, we analyzed the effects before and after the first two-week exposure to messages. The result of the difference-in-differences analysis (Appendix B) shows that users in both Citation and Direct Author groups opened more emails, suggesting habit-forming, but only Direct Author messages significantly boosted the open rates after accounting for the baseline increase in open rates with repeated exposure of alert emails over time (see Fig.~\ref{fig:did-plot} in Appendix B).}
\subsubsection{\new{Importantly, direct author-based messages were significantly rarer}} \label{subsubsection:message_rarity}
While both citation- and direct author-based messages were effective, their prevalence differed significantly. Specifically, direct author-based messages were much \newn{less frequent} ($\mu=0.5$ messages per email, $\bar{\sigma}=1.85$) than citation-based messages ($\mu=1.3, \bar{\sigma}=7.37$, $t(26739.0)=-16.6, p=0$, Welch's two-tailed t-test) (Fig.~\ref{fig:avg_num_featured_papers}) and the pattern remained similar when normalized by the length of emails ($\mu=0.09, \bar{\sigma}=0.212$ in Citation vs. $\mu=0.04, \bar{\sigma}=0.119$ in Direct Author, $t(37428.4)=31.9, p=0$) (Fig.~\ref{fig:avg_frac_featured_papers}).

\new{Yet the frequency of messages was suggested to be a significant factor on \textsc{ctr}. Locally Weighted Scatterplot Smoothing (LOWESS) suggested a significant inverted U-shaped relationship on \textsc{ctr} by the \% of paper recommendations featured in the email (Fig.~\ref{fig:clicked_by_fraction_featured}, Appendix C.). Overall, empirically we found that emails with a greater fraction of treated papers result in dramatically more engagement, up to a point (\newn{approximately} 25--50\% treated) above which engagement falls off. This pattern is conceptually consistent with existing theories of engagement such as the Aristotle's idea of the mean~\cite{grant2011too}, Csikszentmihalyi's optimal difficulty~\cite{csikszentmihalyi1990flow}, and the relationship between workload and innovative work behavior~\cite{montani2020examining}.}

\subsubsection{\new{Optimizing the frequency of direct author-based messages showed a higher estimated marginal utility than for citation-based messages}} \label{subsubsection:study1-frequency}
\begin{table*}[t]
    \centering
    \begin{tabular}{@{\extracolsep{4pt}}c c c c c c c}
    \toprule
    \multirow{3}{*}{Receiver ID} & \multirow{3}{*}{Mail ID} & Dep. Variable & \multicolumn{4}{c}{Predictive Variables} \\
    \cline{3-3}\cline{4-7}
     &  & \multirow{2}{*}{\new{CTR}} & Claimed & \multirow{2}{*}{Receiver h-index} & \multirow{2}{*}{\% Featured} & \# of Total\\
     &  &  & Profile &        &    & Papers\\
    \midrule
    1 & 100 & 1 & 1 & 3 & 0.38 & 14 \\
    2 & 101 & 1 & 0 & 11 & 0.60 & 20 \\
    1 & 102 & 0 & 1 & 3 & 0.30 & 10 \\
    \bottomrule
    \end{tabular}
    \caption{\textbf{Sample format of the \new{collected data used for generalized linear mixed-effects modelling.}} Each row represents a unique user - alert email combination. \new{\textsc{CTR}} is the binary dependent variable \new{representing the email-level user click-through outcome (1: whether any paper recommendation included in an opened email was clicked, or 0: no paper recommendation was clicked)}, \textit{Claimed Profile} shows whether the user has a claimed profile on the search engine, \new{which may indicate an overall high level of engagement.} \textit{Receiver h-index} shows the h-index of the user and was normalized by the avg. h-index of users. \textit{\% Featured} shows the \% of paper recommendations in the email featured with relevance messages. \textit{\# of Total Papers} shows the total number of papers included in the email and was also normalized by the avg. email length in the data corpus before the analysis. \new{Users were randomly assigned to either Control, Citation, or Direct Author conditions, and included as random effects in the model.}}
    \label{table:dataset}
    \vspace{-2em}
\end{table*}
\new{In addition to the \% of paper recommendations featured with relevance messages in the email (\textit{\% Featured}), we also found multiple other covariates suggested to be correlated with user engagement (i.e., \textsc{ctr}) through analysis of descriptive statistics using LOWESS}. For example, users with a higher h-index and a claimed profile had more data available in the system that could be used to feature relevance messages on papers---and these users also tended to show higher baseline engagement (see Fig.~\ref{fig:clicked_by_h} and Fig.~\ref{fig:clicked_by_claimed_profile} in Appendix C). \new{Therefore, we included these covariates as additional predictive variables, and modeled the dependent variable, \textsc{ctr}, using a generalized linear mixed-effects model (GLMM)~\cite{lindstrom1990nonlinear} with the \textsc{lme4} package in R~\cite{bates2014fitting}. GLMMs are often used to analyze (potentially correlated) repeated measures, which in our case corresponds to each user engaging with multiple emails. GLMMs have been used to analyze measurements across many disciplines including medicine, behavioral sciences, and HCI~\cite{cnaan1997using,cudeck1996mixed,hearst2019evaluation,head2021augmenting}.}

\new{We developed increasingly sophisticated models for analysis. For example, our first model (Model 1 in Table~\ref{table:glmm1}, Appendix. E) simply included the \textit{\% Featured} with Direct Author relevant messages and the number of total papers in each email as fixed effects\footnote{Note that though we only report the results from direct author-based relevance messages here, results from the citation-based relevance messages are similar.}.
The result of Model 1 validated the empirical data that showed a curvilinear relationship between \textsc{ctr} and the \textit{\% Featured}.
Our full model (Model 2) added other empirically significant predictive variables described earlier (i.e., \textit{Claimed Profile} and \textit{Receiver h-index}) as fixed effects along with random effects for users to account for user response level correlation (Table~\ref{table:dataset} shows the structure of our dataset).}

\begin{figure}
        \vspace{0pt}
        \centering
        \includegraphics[width=.475\textwidth]{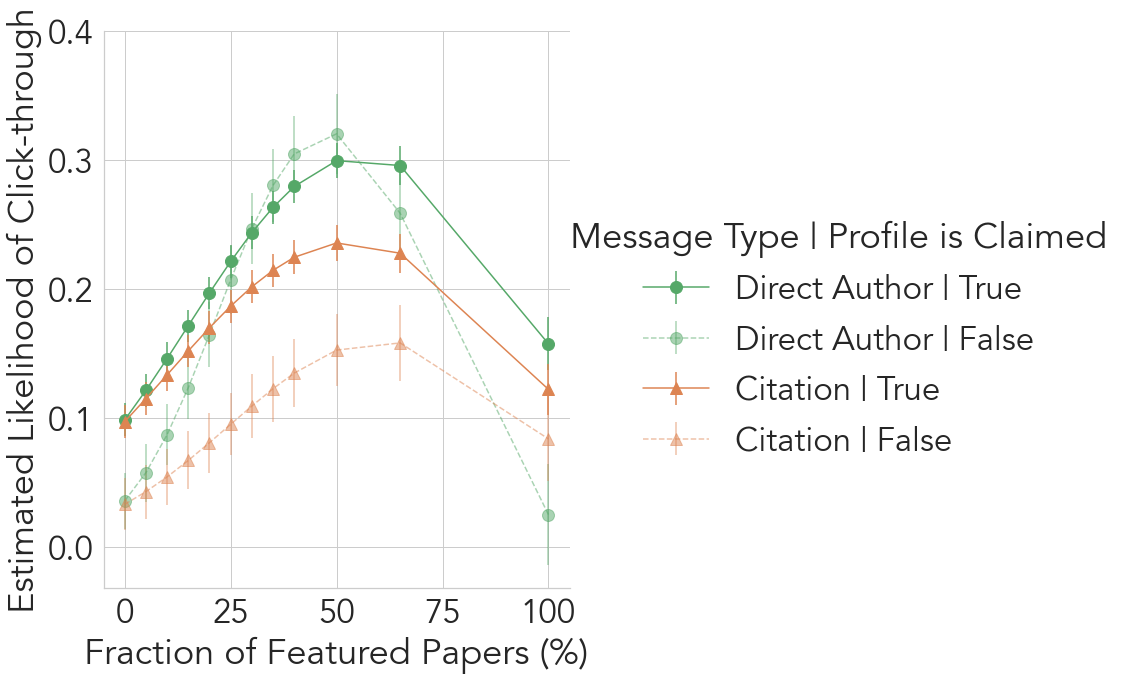}
        \vspace{-1.5em}
        \captionof{figure}{\textbf{Direct author messages showed a significantly higher ceiling of \new{\textsc{ctr}} than citation messages.} \new{The overall estimated (marginal) likelihood of \textsc{ctr} peaked when approximately 50\% of papers were featured with messages, for all groups. However, increasing the \% Featured had a more pronounced effect on \textsc{ctr} for direct author messages (over 20-absolute-percentage-points for both profile-claimed and unclaimed users) than citation messages (15 points or less).}}
        \label{fig:estimated_overall}
        \vspace{-1.5em}
\end{figure}

\new{The full regression result showed once again a significant curvilinear relationship between \textsc{ctr} and the \textit{\% Featured} with direct author-based messages, even after controlling for other covariates (Table~\ref{table:glmm2}). We further estimated the marginal effect of different \textit{\% Featured} for each type of message and user segment representing a high (i.e., user with a claimed profile) versus low level (i.e., without a claimed profile) of engagement. The optimal \textit{\% Featured} was around 50\%, after which the likelihood of \textsc{ctr} dropped off (Fig.~\ref{fig:estimated_overall}). The optimal likelihood of click-through was predicted higher for direct author messages (30\%) compared to citation messages (24\%). The lift from 0-to-optimum \textit{\% Featured} was also predicted higher for direct author messages ($\Delta=+20\%$) compared to citation messages ($\Delta=+14\%$). Taken together,} the analysis suggests that for users with less interaction history and fewer featured papers, strategies to increase the coverage of relevance messages to the 40--60\% range are a promising avenue to increase engagement, \new{and specifically, direct author-based messages may benefit more from \newn{increased coverage compared to} citation-based messages}. We turn to these strategies in Section~\ref{section:indirect-author-message-design}.

\begin{figure}[t]
    \vspace{0pt}
    \centering
    \begin{tabular}{@{\extracolsep{4pt}}r c c c}
    \toprule
     & Coef. & SE & $p$ \\
    \midrule
    (Intercept) & -7.77 & 0.282 & *** \\
    $\text{\% Featured}$ & 4.56 & 0.823 & *** \\
    $\left(\text{\% Featured}\right)^2$ & -4.95 & 1.063 & *** \\
    \# of Total Papers & 0.02 & 0.038 & 0.67 \\
    Claimed Profile & 2.26 & 0.349 & *** \\
    Receiver h-index & 9.53 & 3.476 & ** \\
    $\text{\% Featured}$ $\times$ Claimed Profile & -3.42 & 1.283 & ** \\
    $\left(\text{\% Featured}\right)^2$ $\times$ Claimed Profile & 4.30 & 1.499 & ** \\ 
    $\text{\% Featured}$ $\times$ Receiver h-index & -2.25 & 5.559 & 0.69 \\
    $\left(\text{\% Featured}\right)^2$ $\times$ Receiver h-index & 2.00 & 7.829 & 0.80 \\
    \bottomrule
    \end{tabular}
    \captionof{table}{\new{Regression analysis with our full model (Model 2) predicted a significant curvilinear effect or \text{\% Featured} on \textsc{ctr} in the presence of other covariates (i.e., predictive variables) e.g., whether user has claimed a profile, h-index, and their interactions.} ***: $p<0.001$, **: $p<0.01$.}
    \vspace{-2.5em}
    \label{table:glmm2}
\end{figure}

\begin{figure*}[t!]
    \begin{minipage}[t]{0.24\linewidth}
        \begin{subfigure}[t]{\linewidth}
            \includegraphics[height=0.18\textheight]{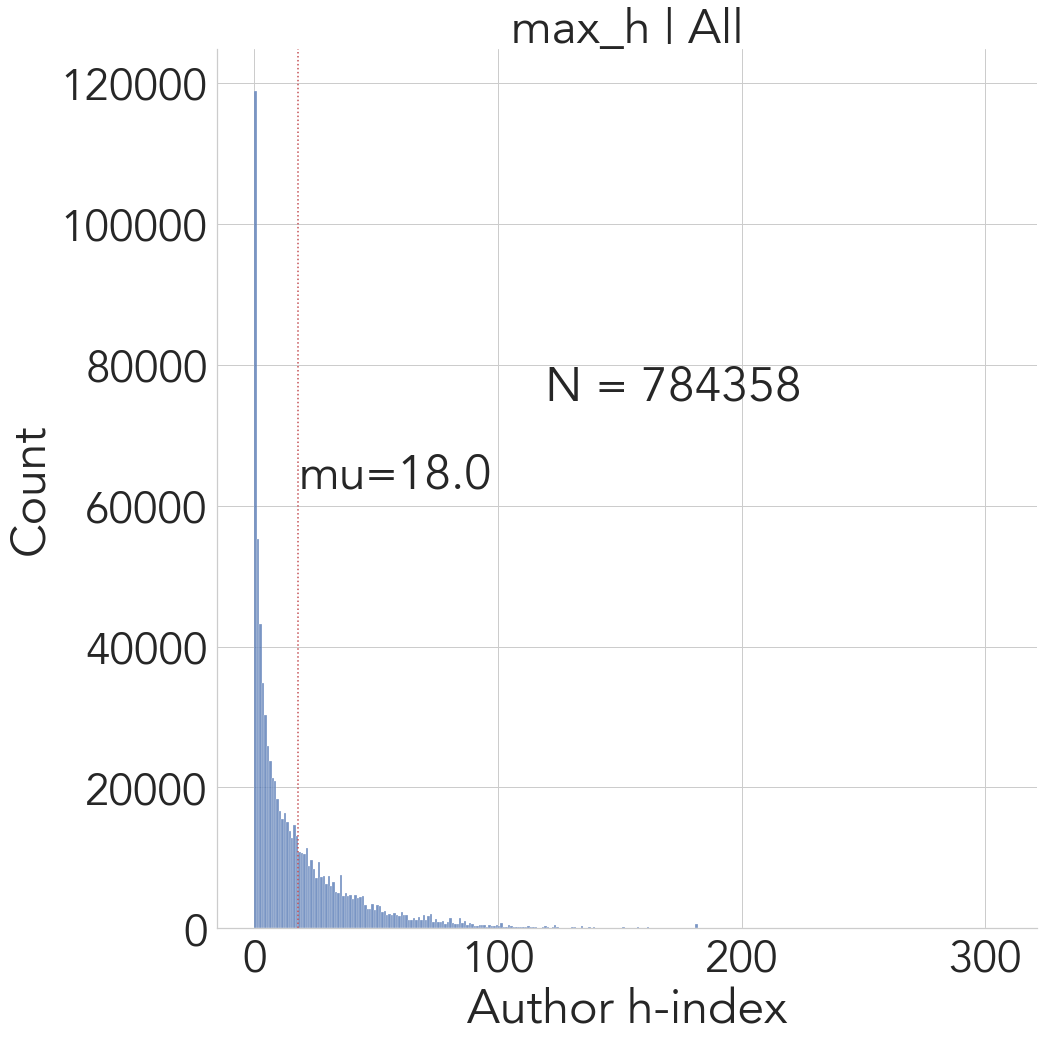}
            \caption{The `background' distribution of maximum author h-index from \new{all paper recommendations sent out in alert emails. The average maximum author h-index was} $\mu=18.0, \bar{\sigma}=23.29$ (median=$9.0$).}
        \end{subfigure}
    \end{minipage}
    \quad
    \begin{minipage}[t]{0.72\linewidth}
        \begin{subfigure}[t]{\linewidth}
            \includegraphics[height=0.18\textheight]{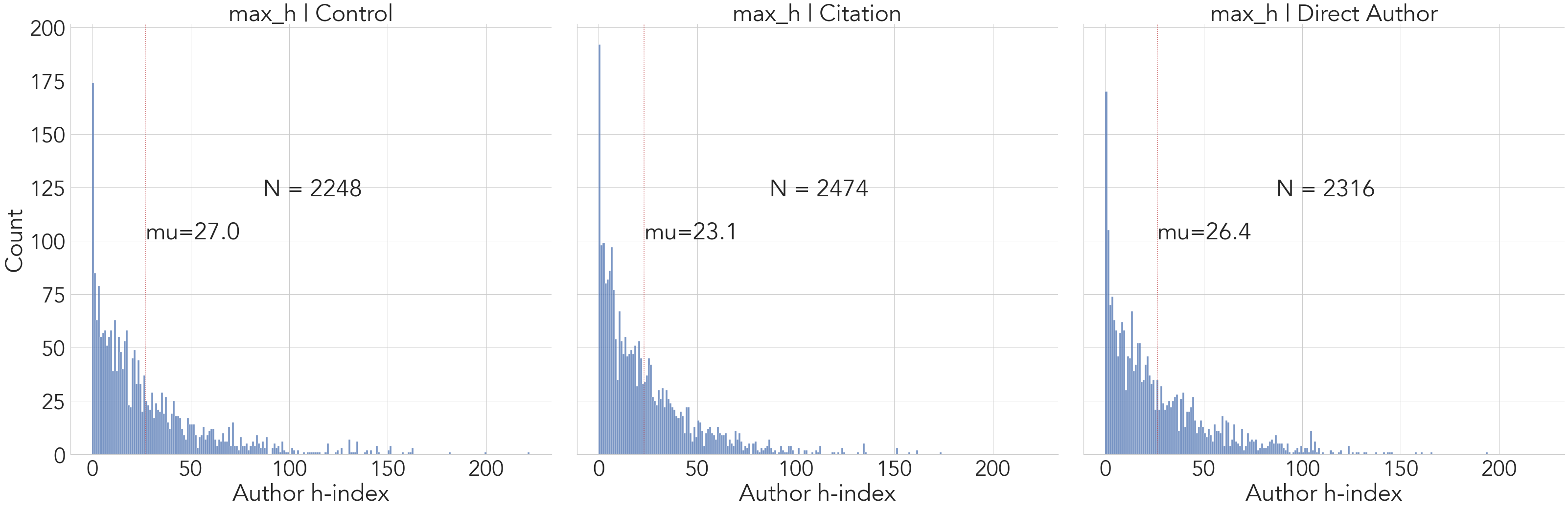}
            \caption{The distribution of maximum author h-index of clicked papers in each condition. The average h-index increased to $\mu=27.0, \bar{\sigma}=29.33$ in Control (median=$17.0$); $\mu=23.1, \bar{\sigma}=24.24$ in Citation (median=$16.0$); and $\mu=26.4, \bar{\sigma}=26.67$ in Direct Author (median=$18.0$). \new{The increase of the average h-index over the background distribution was significant, suggesting that users considered high status authors as a signal for deciding whether to click on a paper by default. At the same time, citation messages reduced the h-index relative to control, suggesting its effect of guiding user attention to lesser known authors. Direct author messages and control did not differ significantly.}}
        \end{subfigure}
    \end{minipage}
    \vspace{-1em}
    \caption{\new{Analysis shows that users clicked on papers featuring high status authors, but does not provide any evidence that messages further centered user attention to them.}}
    \label{fig:h-distribution-max}
    \vspace{-1em}
\end{figure*}
% \begin{figure*}[t!]
%     \subfloat[The `background' distribution of maximum author h-index from \new{all paper recommendations sent out in alert emails. The average maximum author h-index was} $\mu=18.0, \bar{\sigma}=23.29$ (median=$9.0$).]{%
%         \includegraphics[height=0.18\textheight]{figures/all_max_h.png}
%         \label{fig:all_max_h}}
%     \quad
%     \subfloat[The distribution of maximum author h-index of clicked papers in each condition. The average h-index increased to $\mu=27.0, \bar{\sigma}=29.33$ in Control (median=$17.0$); $\mu=23.1, \bar{\sigma}=24.24$ in Citation (median=$16.0$); and $\mu=26.4, \bar{\sigma}=26.67$ in Direct Author (median=$18.0$). \new{The increase of the average h-index over the background distribution was significant, suggesting that users considered high status authors as a signal for deciding whether to click on a paper by default. At the same time, citation messages reduced the h-index relative to control, suggesting its effect of guiding user attention to lesser known authors. Direct author messages and control did not differ significantly.}]{%
%         \includegraphics[height=0.18\textheight]{figures/treatment_max_h.png}
%         \label{fig:clicked_max_h}}
%     \vspace{-1em}
%     \caption{\new{Analysis shows that users clicked on papers featuring high status authors, but does not provide any evidence that messages further centered user attention to them.}}
%     \vspace{-1em}
% \end{figure*}

\subsubsection{\new{Examining the messages' system-wide impact on fairness}} \label{section:h-index-analysis} Though effective, featuring paper recommendations with relevance messages may produce unanticipated negative externalities such as boosting only the visibility of papers by authors who are already often featured in the recommender, \new{or being accessible only by users with high data availability, thus further selectively enhancing their engagement with the literature. Therefore, we examined the fairness of our message designs in two respects: a) their effect on work visibility and b) their coverage for different user groups.}

\xhdr{Fairness of visibility} While the phenomenon of ``rich-get-richer'' has been widely studied~\cite{cole1968visibility,bol2018matthew,merton1968matthew,price1976general,crane1965scientists,judge2007causes,lariviere2010impact,medoff2006evidence,diprete2006cumulative,van2014field,wang2014unpacking}, \new{to our knowledge no study has examined the effect of interventions designed to shift user attention to certain types of papers on visibility via a randomized study on a large-scale, real-world, deployed recommender system.}

One way to examine relevance messages' effect on visibility is to take an outcome oriented approach, by measuring changes on the papers users clicked before and after intervention. In our alert emails when a user clicks on a paper link, it brings the user to an interactive paper detail page, which includes full abstract, the link to a full-text file, any figures, and the author information among others. Therefore the click interaction serves as a strong indicator of the user's exposure to the paper's content. Thus we operationalized the visibility of each paper as a binary measure of whether it was clicked; when a paper is clicked for the first time, the paper webpage is made visible to the user. Using this measure of visibility, \new{we investigated whether its distribution over the academic status of the authors of papers changed systematically} with the introduction of relevance messages. 

To operationalize the academic status of each author, we used the h-index measure. The index, since its introduction by Hirsch in 2005~\cite{hirsch2005index}, has been popularized as a metric of academic success. Though limitations of the index exist~\cite{panaretos2009assessing,sinatra2016quantifying}, it is seen as a robust measure~\cite{hirsch2007does,radicchi2008universality,henzinger2010stability,acuna2012predicting}, and informs high stakes decisions such as hiring, promotion, and funding~\cite{mcnutt2014measure,hicks2015bibliometrics,abbott2010metrics}. It is also widely featured in many scholarly search engines and citation databases. We assigned an h-index to the paper recommendation by taking the max h-index of authors on the paper. \new{This is plausible because the highest status author of the paper may appear salient and easily recognizable to the user at first glance.} % and aggregated it over all recommendations sent out in alert emails.
This produced our baseline h-index distribution of authors (Fig.~\ref{fig:h-distribution-max}, first panel on the left). Next, we similarly computed the h-index of each recommendation but using only the clicked paper recommendations from each condition (Fig.~\ref{fig:h-distribution-max}, three panels on the right). Finally, \new{to examine whether relevance messages led to users clicking papers with high status authors more often}, we tested whether the average h-index of clicked papers significantly differed from that of the baseline.

The result shows that
% \new{while featuring papers with relevance messages significantly increased the h-index of authors of the clicked papers, the increase was no worse than the baseline,}
\newn{while the h-index of clicked papers was significantly higher than the h-index of all papers, this increase was no worse than the baseline,}
suggesting that users naturally incorporate author identities and \new{status represented as} h-indices when deciding whether to click on a paper recommendation (\new{shown in the jump from all authors' h-index to clicked authors' h-index} $\mu=18.0 \rightarrow \mu=27.0$; Fig.~\ref{fig:h-distribution-max}, first two panels on the left, $t(2255.13)=-14.55, p=0$, Welch's two-tailed t-test), \new{and featuring relevance messages did not} \newn{exacerbate this effect}. \new{In addition, there was evidence that} citation \new{messages guided user attention towards lower status authors' papers more than Control} ($t(4371.48)=5.06, p=4.36\times10^{-7}$). \new{For robustness against any spurious effects from choosing the maximum h-index, we repeated the analysis using the average h-index of authors and found that the patterns remained the same (see Appendix F).} Taken together, we conclude that augmenting paper recommendations did not \newn{adversely impact} fairness of visibility with respect to h-index, and may have \new{shifted towards a fairer distribution of user attention when} citation-based messages were used.
\begin{figure}
    \begin{center}
        \includegraphics[height=0.22\textheight]{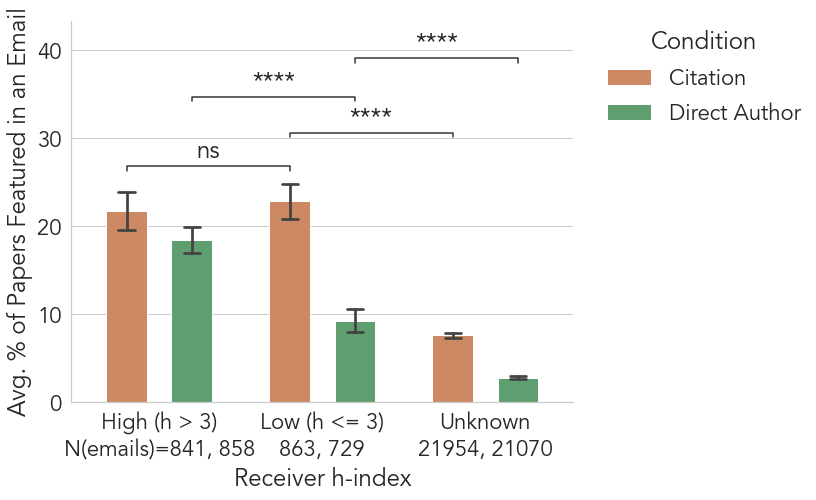}
        \vspace{-1em}
        \caption{While the frequency of direct author messages decreased significantly, from 18\% to 9\%, as users' h-index decreased from High to Low, it remained similar for Citation.}
        \label{fig:avg_frac_featured_papers_h_group}
    \end{center}
    \vspace{-2em}
\end{figure}

\xhdr{Fairness of coverage} \new{While Direct Author messages were rarer than citation messages overall (Fig.~\ref{fig:avg_frac_featured_papers}), was there any systematic difference in the coverage of user groups with varying academic status?} We further divided the users into three groups of h-index (Low, High, and Unknown) using the median (3) h-index from the authors whose h-indices were available (h-index was known for 115/{2,474} users in the Citation condition and 107/{2,316} users in the Direct Author condition). \new{We found that groups received significantly fewer Direct Author messages as the h-index got smaller: $\mu=18\%$ in the high h-index group, $\mu=9\%$ in the low h-index group, and $\mu=3\%$ for the group with unknown h-index (the pairwise decreases were significant, $p<.0001$, Fig.~\ref{fig:avg_frac_featured_papers_h_group}). This may not be surprising given a user's high h-index may be correlated with an overall higher level of engagement with the search engine that results in richer interaction data and also higher connectedness within the academic network (e.g., a bigger collaborators' network), useful for message generation.} However, the frequency of citation-based relevance messages did not significantly differ between the Low and High index groups, suggesting a relative abundance of citation-based relations for users with fewer publications \new{(e.g., our citation message generator could find relevance through \newn{lower index} users' curated feeds alone)}; \newn{however, the same was not true for author-based messages}.
% \new{This result suggests that fairer author-based relevance messages involve increasing the coverage for less connected users in the academic social network.}
\newn{This result suggests that increasing coverage of messages may be particularly helpful for improving the fairness of author-based relevance messages, and benefit users who have fewer connections in the academic social network.}

\section{Design of Indirect Author-based Relevance Messages} \label{section:indirect-author-message-design}
\subsection{\new{Key Motivations and Related Work}}
\new{The key motivating results of Study 1 showed that 1) Direct author-based messages were an effective mechanism for increasing future engagement (the overall email open rates were significantly higher in the direct author-based messages condition compared to Control and the citation-based messages condition (Fig.~\ref{fig:open_rates}), and the effect was clear after controlling for the baseline effect from habit-forming (the difference-in-differences analysis, Appendix B); 2) This effectiveness was achieved despite the direct author-based messages being significantly rarer than citation-based messages (Section~\ref{subsubsection:message_rarity}). This relationship may generalize, possibly enabling higher \textsc{ctr} with fewer messages/email. For example, in our GLMM analysis, when the likelihood estimate was held as constant, the estimated \% Featured is lower for direct author-based messages than citation-based messages (Fig.~\ref{fig:estimated_overall}; 3) In addition, direct author-based messages showed a higher ceiling of the (marginal) likelihood of click-through compared to citation-based messages (Fig.~\ref{fig:estimated_overall}). Thus, in order to achieve the author-based relevance messages' full potential, we designed a novel mechanism for expanding the coverage by incorporating indirect relations mediated by trusted intermediate authors between a user and the recommended papers.}
%, and finally 4) the relatively limited coverage for low h-index users, due partly to limited data and few direct connections found in the academic social network which is problematic.

\new{Prior work in social network-centric relevance explanations further provides support for expanding author messages via indirect relations. %(a) they are likely to be effective and (b)  will inform open questions about their design and application to the scientific domain.
For example, Sharma and Cosley investigated the value of featuring \textit{direct} friendship-based relevance messages on persuasiveness and informativeness in the music recommendation domain~\cite{sharma2013social}. They found relevance through good friends mattered more than random friends, and as such showing the (good) friends' names in messages led to higher informativeness than representing them as aggregate friend popularity (e.g., `3 of your friends liked this album'). Additionally, when messages featured the overall popularity (e.g., `{12,211} of Facebook users like this'), the popularity mattered only if the users identified with the crowd. Extrapolating these findings to the design of indirect author messages, we expect to find similarities such as the importance of relation strength and specifying which author and why they were featured in the messages, but also new challenges as to deciding who should be the intermediate authors and how to identify them. We expect the citation network of existing publications and their authors to contain topical relevance and trust relations targeted specifically to our task domain, and this informs our algorithm design for \textit{inferring} an academic `social' network from it. In this regard, our approach differs from~\cite{sharma2013social,sharma2011network} which used existing social networks (e.g., Facebook) and~\cite{chen2010short} which relied on Twitter. Additionally, the SONAR system combined social information for members within an organization from co-authorships of organizational Wiki articles and user interaction traces such as bookmarking the same pages and usage of same tags~\cite{guy2009personalized,guy2008harvesting}. However, this work also differs from our work due to its dependence on an explicit social network and its limited scope to direct relations. In the subsequent sections, we describe the design of indirect author-based messages and its generation algorithm.}

% From the success and limitations of direct author-based relevance messages in Study 1, we were motivated to design an improved version that are likely to retain relevant benefits while reducing its limitations. Specifically, we aimed at expanding the access and coverage of author-based relevance messages on paper recommendations, the primary limitation found from the results of Study 1. Among possible alternative approaches for achieving this goal, we chose the mechanism of indirect author-based relations, in order to meet the following three design goals.

% Motivated by prior work ... Sharma and Cosley ...; Practical issues ... finding trusted authors and receiving recommendations from them, e.g., on Twitter... mention it in Related Work and refer back to it?
% The SONAR system infers social information within an organization by aggregating organizational chart relationships, explicit connections originating from enterprise social networks, and co-autxhorships of organizational Wiki articles etc.~\cite{guy2009personalized,guy2008harvesting}. While the scholarly recommendation domain lacks explicit network information used in the SONAR system, the inference of relationships from aggregation of co-authorship information can provide an adaptable design for the domain.

\subsection{Design Goals}
\new{We grounded the design of indirect author-based messages with the following goals:}

G1. \textit{\new{Support relevant information needs}.} Scientists often turn to trusted sources as a way to curate relevant papers. Yet, this process is often tedious and existing tools provide piecemeal support at best. Therefore, messages should provide benefits similar to receiving personalized bibliography from an expert source. % based on prior research on scholarly information behaviors that reported current practices and pain points experienced by researchers~\cite{radford2020people,megwalu2015academic} % Researchers often follow email discussion threads, use colleagues as information sources, or `cold email' high-profile experts in an outside field to receive highly valuable bibliography~\cite{vakkari2006searching,covi1999material,palmer2005scholarly,palmer2009scholarly}. However, these solutions are still piecemeal and tedious, and relied upon chance discovery~\cite{breitinger2019too}.  It was also motivated by prior work in computational approaches for recommending content based on the social network of trusted sources~\cite{chen2010short,andersen2008trust}

G2. \textit{Support serendipitous discovery,} \new{which is an oft-mentioned benefit~\cite{radford2020people,megwalu2015academic,breitinger2019too}. To achieve this, the author featured in messages should be likely subjects of serendipitous discovery.} %Because we expect the likelihood of an author being known to the user to inversely correlate with their academic status (i.e., h-index), we filter triplets that exceed the maximum h-index constraint for [author].

G3. \textit{Support design continuity.} \new{New message designs should leverage the robust designs used in Study 1, to minimize unanticipated negative consequences from design changes, and to prevent harming the user experience of the platform.}
% One of the main reasons why we based our design off of the author-focused relevance message was the success from Study 1. Our design should leverage the robust and tested message designs used in Study 1 whenever possible to minimize the chance of any unanticipated negative effects from design. Therefore, we borrow the overall format and visual elements from the earlier message and minimally vary the textual information when appropriate for the new design (fig.~\ref{fig:indirect_author_combined}, left). Because there are two author names associated with each message, we annotate an asterisk next to the author's name to differentiate from the indirect author. We also added an underline and click interaction to the design, which was missing in the messages featured in Study 1 yet was regarded valuable from informal interviews with users. We also added a second-line text to the message to explain the user's relation to the indirect author (``You annotated...'') for clarity.

\begin{figure*}[t]
    \includegraphics[width=\textwidth]{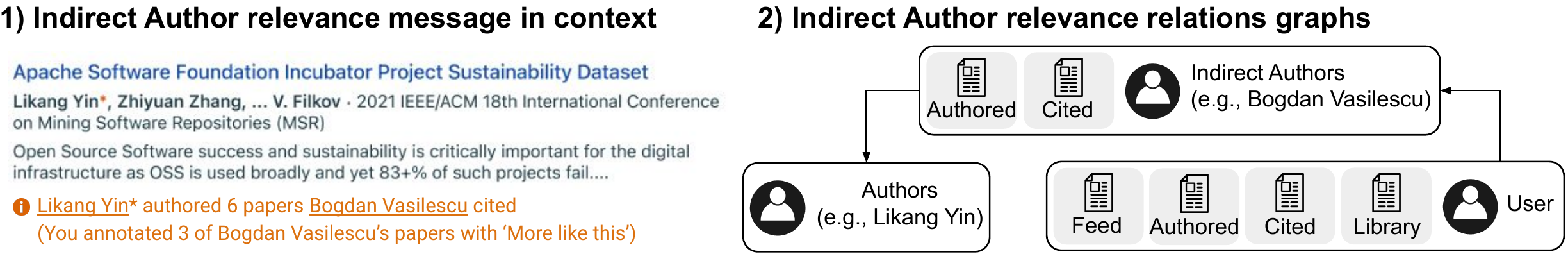}
    \caption{\textbf{Indirect author relevance message design.} \new{1) A sample message rendered in the recommendation context as shown to the user. The message features two lines of text. The first line features the relation between the author of the paper (Likang Yin, underlined) and the indirect author (Bogdan Vasilescu). In the second line, the user's relation to the indirect author is described. The author names are clikable and linked to profile pages on the search engine. 2) Indirect author candidates are first identified from a set of user papers and interaction data then filtered based on their connection to the authors of the paper.}}
    \label{fig:indirect_author_combined}
    \vspace{-1em}
\end{figure*}

\subsection{Message \new{Implementation and} Generation Algorithm}
\new{We adopted the overall textual and visual design from earlier message designs, with an additional line of text to accommodate the relational information (Fig.~\ref{fig:indirect_author_combined}, left). We placed an asterisk next to the author's name to differentiate it from the indirect author's. We also added underlines and click interaction to the names to support a discovery experience. The second-line text was added to describe the user's relation to the indirect author for clarity.}

\new{To generate the messages, candidate intermediate \newn{(indirect)} authors were identified from the user's publication and interaction data. Then, each candidate's publication data was used to extract [author]-[indirect author]-[user] triplets for each author of a recommended paper. Finally the first ranked triplet was fed into the message template (Fig.~\ref{fig:indirect_author_combined}, left).}

To rank the triplets, let us first define the strength of the [author]-[indirect author] relation as `Relevance' and the strength of the [indirect author]-[user] relation as `Influence'. For a candidate $(\text{author}, \text{indirect author}, \text{user}) = (i, j, u)$ triplet, we computed these strengths as
\begin{align*}
    \text{Relevance}_{i,j} &:= \new{a}\times\text{log}(\newn{\text{co-authored}_{i,j}} + 1) + \new{b}\times\text{log}(\newn{\text{cited}_{i,j}} + 1) \\
    \text{Influence}_{j,u} &:= \text{log}(\newn{\text{engaged}_{j,u}}) \times \left(j\text{'s h-index}\right),
\end{align*}
\new{\newn{where $\text{co-authored}_{i,j}$ is the number of papers that $i$ and $j$ co-authored, $\text{cited}_{i,j}$ is the number of $i$'s papers $j$ cited, $\text{engaged}_{j,u}$ is the number of $j$'s papers $u$ engaged with using one of the following actions: co-authoring, citing, saving, and annotating with `more like this' on a personal research feed.} The intuition here is that the higher the number of actions $i$ took on $j$'s papers (e.g., citing, saving), the stronger the tie strength. The constants $a$ and $b$ in Relevance control the relative strength of relations, and we set $a=2, b=1$} to weight co-authorship twice as strong as citation \new{because the former is believed to indicate stronger relevance.} In addition, we take a logarithm of the count of papers to account for the diminishing signal strength (e.g., an extra citation means much less when it is already cited 20 times). For Influence, we multiply the logarithm of \newn{engagement} counts with the candidate indirect author's h-index to prioritize individuals with higher academic status because they are more likely to be known and trusted by the user. Finally, our ranking objective \newn{for a given user $u$ is:} $\text{argmax}_{i,j}\left(\text{Relevance}_{i,j} \times \text{Influence}_{j,u}\right)$. This multiplicative objective was designed to prioritize triplets with coherent (rather than lopsided) tie strengths (e.g., a high score on only one of Influence or Relevance but low score on the other may result in an overall irrelevant relation to the user).
%We set $\text{Relevance}_{ji} = \text{Relevance}_{ij}$ and $\text{Influence}_{ji} = \text{Influence}_{ij}$.

\section{Study 2 - Controlled Lab Study} \label{section:study2}
We performed a formal usability study to gain insights into the following questions: How do different types of relevance messages aid scientists' ability to review the recommended research papers in an email alert context? How do scientists make sense of the relevance information conveyed in the messages? What are the challenges and design implications for future author-based relevance messages?

Using a within-participants design, we compared the Indirect Author-based relevance messages to Citation- and Direct Author-based relevance messages and Control (no message). The quantitative and qualitative results were in favor of the relevance messages, and the value of different types of messages seemed complementary. Through open coding of interview transcripts, we discovered different themes of the benefits complementing the results from Study 1. We also uncovered challenges from which we synthesized implications for design. 

\subsection{Study Design}
\xhdr{Participants} 14 scientists were recruited via university and company mailing lists. 1 was an assistant professor, 5 were postdoctoral researchers, 1 was a professional researcher, 1 was a Master's student, and 6 were doctoral students. 6 of the 14 participants identified their discipline as human-computer interaction. Participants were compensated at a \$30/hour (USD) rate. The study sessions were between 1-hour- and 1.5-hours-long and held remotely on a video conferencing platform. Participants opened an individualized Google Doc prepared by the interviewer and were asked to share their screen. After obtaining consent from each participant, the interviewer proceeded to record the session.
%The recordings were later transcribed and analyzed.

\xhdr{Stimulus recommendation emails} Personalized paper recommendations were generated for participants using their publication, library, and research feed data. Each participant's recommendations were randomly subdivided into 4 sets (A, B, C, D) of equal length, ranging from 12 to 30 papers per set. Then, %In B, C, and D recommendations, citation-, direct author-, and indirect author-based 
relevance messages were generated and added to the corresponding paper recommendations. Following the results from Study 1 (Section~\ref{subsubsection:study1-frequency}), we allowed up to 50\% of the papers in each email to be featured, which was the approximate optimal fraction of papers to be featured. The generated emails looked exactly the same as in Study 1, except for the messages and the headers, which were anonymized so as to not include any identifying information when recording with screen share. %\new{We controlled the visual appearance of emails across the conditions to better compare the user experience between conditions.} 
In Control (A), paper recommendations were shown without messages. In the citation- (B), direct author-based (C), and indirect author-based (D) conditions, messages of the corresponding type were added underneath each paper recommendation, when applicable.

% \begin{wraptable}{R}{.3\textwidth}
% \begin{minipage}{.3\textwidth}
%     \includegraphics[width=\textwidth]{figures/study2_overall_helpfulness2.png}
% \end{minipage}
% \begin{minipage}{.3\textwidth}
%     \includegraphics[width=\textwidth]{figures/study2_message_helpfulness2.png}
% \end{minipage}
% \begin{minipage}{.3\textwidth}
%     \includegraphics[width=\textwidth]{figures/study2_message_information2.png}
% \end{minipage}
% \begin{minipage}{.3\textwidth}
%     \includegraphics[width=\textwidth]{figures/study2_ease2.png}
% \end{minipage}
% \begin{minipage}{.3\textwidth}
%     \includegraphics[width=\textwidth]{figures/study2_second_line_helpfulness2.png}
% \end{minipage}
\begin{figure}
    \vspace{-1em}
    \includegraphics[width=.27\textwidth]{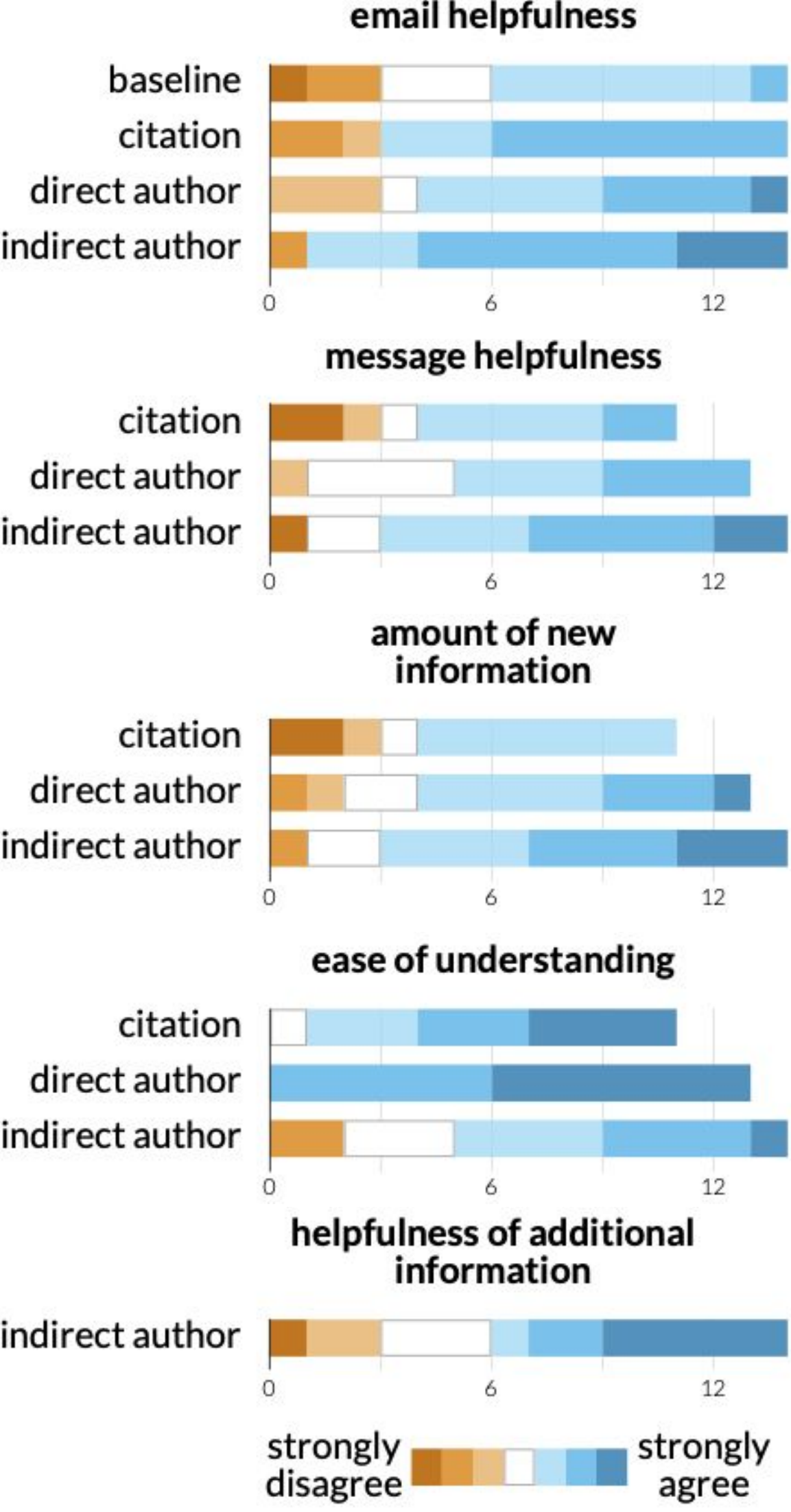}
    \caption{\textbf{Subjective responses to test questions}. \new{Overall, participants responded favorably on the general helpfulness of messages, with indirect author messages being harder to understand. The benefits of different message types seemed complementary (see text).}}
    \label{table:likert}
    \vspace{-2em}
\end{figure}

%\xhdr{Recommendation emails} In Control (A), paper recommendations were shown without messages. In citation- (B), direct author-based (C), and indirect author-based (D) conditions, messages of the corresponding type were added underneath each paper recommendation, when applicable.

\xhdr{Tasks and Assignment} Each session ran as follows: 1) Greeting and obtaining consent for recording; 2) Background questions around how the participant typically obtains research papers to read. If the participant was a user of an alert system, the interviewer also asked questions about their experience with typical new paper alerts; 3) Complete four timed tasks (6 minutes each), each of which was followed by a task-specific questionnaire. Using four 4$\times$4 Latin Square blocks, we assigned each participant to one of the randomly drawn rows. The number of recruited participants (14) resulted in a near but not fully factorial design (two presentation orders had one more participant each).

\xhdr{Measures} For each task, we measured the following (all but the last two measures are on a 7-point Likert scale between 1: Strongly disagree and 7: Strongly agree). In 4 condition blocks (3 in B), participants skipped measures related to relevance messages, as no messages were shown to them. \new{``\textit{Email helpfulness}'' is the participant's self-assessed agreement with the following statement: ``I found the email helpful.''; ``\textit{Message helpfulness}'' (in B, C, D) indicates the participant's self-assessed agreement with the following statement: ``I found the orange text underneath paper recommendations helpful.''; ``\textit{Novel information}'' (in B, C, D) indicates the participant's self-assessed agreement with the following statement: ``I found the orange text underneath paper recommendations to contain interesting new information.''; ``\textit{Ease}'' (in B, C, D) indicates the participant's self-assessed agreement with the following statement: ``It was easy to understand what the orange text underneath paper recommendations was trying to tell me.''; ``\textit{Additional information helpfulness}'' (in D) indicates the participant's self-assessed agreement with the following statement: ``I found the second line of the orange text helpful.''; ``\textit{\% Featured}'' (in B, C, D) is the percentage of paper recommendations featured with relevance messages; ``\textit{Number of familiar author/indirect author names}'' (in D) is the number of author/indirect author names included in indirect author-based relevance messages the user found familiar.}

We distinguish between the overall email helpfulness question and the message type-specific questions for two reasons: one, the overall helpfulness measure allows us to compare the usefulness of message-augmented emails with that of the baseline; two, message type-specific measures allow us to see whether participants saw complementary value from the different types of messages, in which case we expect to see no significant difference between \newn{the message} conditions. 

\xhdr{Analysis} For each of the quantitative measures, we ran post-hoc analyses with Welch's two-tailed t-test. For qualitative analysis, the interviews were recorded, transcribed, and coded in four iterations following an open coding approach. The goal of this process was to identify common themes that captured rich qualitative insights grounded in data~\cite{boyatzis1998transforming}. Using the transcripts, audio, and video recordings of the interviews, two coders first independently performed open coding of the initial themes. Subsequently, they had in-depth discussions over multiple sessions to merge similar codes and form higher level themes, as well as a larger group discussion with four of the authors on the themes. Finally, these consolidated codes and themes were applied to the transcripts to arrive at the qualitative findings of Study 2. We focused our analysis on uncovering the benefits and challenges of relevance messages, and the results converged on themes presented in Table~\ref{table:qualitative_overview}. The detailed descriptions of these themes are provided in Section~\ref{subsection:qualitative_results}.

\subsection{Quantitative Results}
\begin{figure}
    \begin{minipage}{.225\textwidth}
        \centering
        \vspace{-1em}
        \includegraphics[height=0.21\textheight]{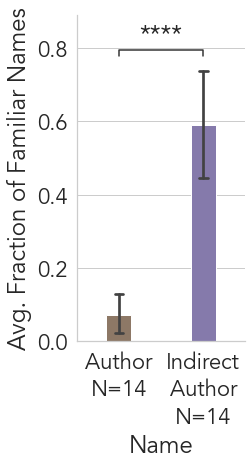}
        \vspace{-1em}
        \caption{\new{On average, participants recognized indirect author names about 60\% of the time, and author names 7\% of the time (8 times less than indirect authors).}}
        \vspace{-1.5em}
        \label{fig:familiar_names}
    \end{minipage}
    \quad  
    \begin{minipage}{.225\textwidth}
        \centering
        \vspace{-1em}
        \includegraphics[height=0.21\textheight]{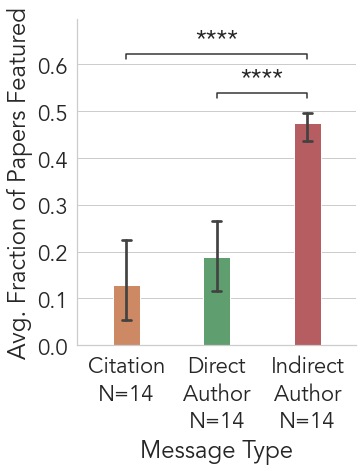}
        \vspace{-1em}
        \caption{\new{Indirect author messages significantly increased the coverage to 47\%, compared to direct author (13\%) and citation messages (19\%).}}
        \vspace{-1.5em}
        \label{fig:num_messages}
    \end{minipage}
\end{figure}
The result \new{validated the feasibility of indirect author messages}. The \% of papers featured with indirect author messages was close to the estimated optimal, at 47\% ($\bar{\sigma}=0.062$). \new{In comparison,} citation- ($\mu=0.13, \bar{\sigma}=0.175$) and direct author-based messages ($\mu=0.19, \bar{\sigma}=0.156$) \new{were significantly less frequent} (at $p<0.00001$, Welch's two-tailed t-test, Fig.~\ref{fig:familiar_names}). \new{In addition, indirect authors were significantly more recognizable than the authors} ($\mu=0.59, \bar{\sigma}=0.290$ vs. ($\mu=0.07, \bar{\sigma}=0.111$) ($t(16.74)=-6.25, p=9.4\times10^{-6}$, Fig.~\ref{fig:familiar_names}).

\new{Overall, participants perceived indirect author messages favorably. The result of a post-hoc t-test showed that emails augmented with indirect author messages ($\mu=5.7, \bar{\sigma}=1.27$) were perceived as significantly more helpful than the baseline ($\mu=4.1, \bar{\sigma}=1.46$) ($t(25.49)=-3.04, p=0.005$), but for citation- ($\mu=4.8, \bar{\sigma}=1.66$) and direct author-based messages ($\mu=4.9, \bar{\sigma}=1.32$) the difference was not significant (Fig.~\ref{table:likert}, top). At the same time, participants found indirect author messages the most difficult to understand (Fig.~\ref{table:likert}, second from the bottom, avg. response was $\mu=4.9, \bar{\sigma}=1.21$ for indirect author messages and $\mu=6.5, \bar{\sigma}=0.52$ for direct author messages, $t(17.92)=4.56, p=0.0002$). Six participants (42\%) also responded either in a negative (21\%) or neutral (21\%) manner to the additional text designed to help with understanding (Fig.~\ref{table:likert}, bottom). Thus, the message usefulness may have been reduced by the difficulty of understanding. Furthermore, we hypothesized that each message type may bring complementary value to the user when they engage with paper recommendations. For both \textit{Message helpfulness} ($p>0.14$) and \textit{Novel information} ($p>0.18$) questions, we found no evidence of significance differences in user responses. P14 commented that ``each of these messages feels like it has a different utility than the other two.'' We further investigate this hypothesis with qualitative analysis of the data below.}

\subsection{Qualitative Results} \label{subsection:qualitative_results}
\begin{table*}[t]
    \centering
    \begin{tabular}{p{1.25cm} p{6cm} p{9.25cm}}
    \toprule
    \multicolumn{3}{c}{\textbf{Benefits and Challenges of Relevance Messages Augmenting Scientific Recommendations}} \\
    \midrule
    Benefit & B1: Drawing Attention & Design and personalization of messages were sufficient to draw user attention\\
    \addlinespace[.1cm]
    Benefit & B2: Social & \new{Developing} awareness of other scientists \\ 
    \addlinespace[.1cm]
    {Benefit} & B3: Discovery of Community-Related Insights & \\
    \addlinespace[.1cm]
    & \textit{B3-a: Connections in the Intellectual Community} & \new{Seeing} connections in the community \\
    \addlinespace[.1cm]
    & \textit{B3-b: Scientist's Intellectual Impact} & \new{Understanding \& tracing} their impact\\ 
    \addlinespace[.1cm]
    Benefit & B4: Mobilizing Mental Models & \\
    \addlinespace[.1cm]
    & \textit{B4-a: Accessing Readily Available Mental Models} & \new{Understanding} the recommended research \\
    \addlinespace[.1cm]
    & \textit{B4-b: Applying Mental Models for Transfer} & \new{Making} sense of connections to \new{new} authors \\ 
    \addlinespace[.1cm]
    Benefit & B5: Serendipitous Discovery & \new{Discovering} new and interesting authors \\
    \addlinespace[.1cm]
    Benefit & B6: Judging Potential Value & \new{Judging} the usefulness and interestingness of recommended papers \\ 
    \addlinespace[.1cm]
    \midrule
    \addlinespace[.1cm]
    Challenge & C1: Interpretability & Sources of misinterpretation: linguistic and semantic mismatches, incongruence between recommendation and relevance messages \\ 
    \addlinespace[.1cm]
    Challenge & C2: Context vs. Efficiency & Tension between wanting to see more contextual information and efficiency of reading \\ 
    \addlinespace[.1cm]
    Challenge & C3: Scientist's Evolving Identity & Challenges from how scientists' interests change over time and move away from old topics \\ 
    \addlinespace[.1cm]
    Challenge & C4: Trust & Challenges due to errors in relevance quantification and increased user sensitivity \\
    \bottomrule
    \end{tabular}
    \caption{\textbf{Different types of benefits and challenges of relevance messages augmenting paper recommendations.}}
    \label{table:qualitative_overview}
    \vspace{-2em}
\end{table*}

Now we present qualitative insights into participants' experience with messages. Using axial coding following our open coding, we organized these insights into six types of benefits and four types of challenges (Table~\ref{table:qualitative_overview}) and provide detailed descriptions below. \new{We use a fraction (e.g., 3/14 participants) to represent multiple participants expressing similar thoughts. In participant quotes, we used notations \textbf{[A]} and \textbf{[IA]} to denote \textbf{Author} and \textbf{Indirect Author} names featured in the message, respectively.}
%These findings complement earlier quantitative results on the effectiveness of relevance messages seen in Study 1. 
% Intro mentions `on' vs. `around' benefits --- should this be another categorization in the table?

\xhdr{Types of Benefits} \textit{B1: Noticing and being drawn into relevance messages.} Participants in the interviews commented that the overall design of relevance messages stood out to draw their attention to the recommended papers featured with messages: ``It seems to be all the orange text sticks out... So that's why I would scroll through the email and really look out for those'' (P9). Participants mentioned that they `liked' the visual distinction of the message design \new{(3/14 participants)}, they `missed' the messages when there was none after moving to the next task in the experiment \new{(P2)}, they used them as anchor points to structure their experience and attention around relevant items \new{(4/14 participants)}, and that they were drawn to the personalized connections described in the messages (P1). 
%``It seems to be all the orange text sticks out -- when it's orange, or red, for some reason, I feel like it's there to draw your attention to sort of like a warning almost. Especially with this, like, (i)-symbol, it just seems like the system wants you to go there first, some higher importance. So that's why I would scroll through the email and really look out for those'' (P9);
% ``I kind of do miss the orange text now'' (P2); ``The `you' part attracts my attention, and I really notice how I have something connected to this paper; It's not something magical that your prediction network suggested this but there is a real relation'' (P1). These quotes suggest that messages were indeed easily noticed by the users and shifted their attention towards recommendations featuring them.

\textit{B2: Developing awareness of other scientists.} Participants were curious and wanted to see work by other scientists they cared about to gain a general understanding of what their most recent research areas were: \new{``Oh, this is what [A] is doing... So now I can be like, `Hey, I saw the UMAP paper you published' when I meet him in two weeks. So a conversation point'' (P4).} Author-based relevance messages were appreciated because they surfaced author connections that might otherwise have been unnoticed or easily missed: ``This paper is not something interesting to me but the personal connection is. I'm also surprised that my PhD advisor highly cited papers by him. So maybe I should have a look at it more closely?'' (P3).
%``okay it is [Author]! So this paper is not something interesting to me but the personal connection is. He was working in my department some months ago... I'm also surprised that my PhD advisor highly cited papers by him. So maybe I should have a look at it more closely? I did not think that [Author] would have written articles that I would be potentially interested in, so that's what got me curious. But it's not the article that's really interesting but that you can see oh this person does things in my community which was unexpected'' (P3); ``it's like, sometimes cool to be like, Oh, this is what [Author]'s doing... So now I can be like, Hey, I saw the UMAP paper you published when I meet him in two weeks. So a conversation point for this'' (P4). 
In such cases, even divergent topical areas did not seem to deter participants' curiosity:
%``I'll probably open it just to know what he's working on.'' (P14);
``Not directly relevant to my research but I would probably look at this because I know the author'' (P9); ``I don't cite \new{[A]} anymore. It's more like, just curious what he's up to'' (P4).
%When participants had a deeper understanding or interest in the work by an author represented in the message, it was valuable to get an updated view on how their research direction was evolving
\new{Messages were valuable as a vehicle for getting updated information on how the research direction of an author represented in them was evolving:} ``I know \new{[A]}. I would like to work with her. So maybe I'm going to read this paper because just to know what she's up to'' (P6); 
``interesting that \new{[IA]} is working on voice interaction too'' (P7).
%``[Author] authored three papers [Indirect Author] cited... interesting that [Indirect Author] is working on voice interaction too'' (P7).

\textit{B3: Discovery of Community-Related Insights.} Messages also helped participants recognize community-related connections that \new{go beyond individual author connections} in two ways: \new{first, messages} helped them make sense of their connections with broader research communities; \new{second, messages} helped them understand the impact of their work.

\textit{B3-a. Connections in the Intellectual Community.} Participants found author-based relevance messages helpful for understanding how they might be connected to larger research communities: ``Interesting... because at least in my community it feels a bit like a small world. So it's interesting to know who's cited by who'' (P3); ``Okay so [IA] and [A] have a deep connection in the research space and I have a weak connection with this space with them'' (P6); ``It's like, `Hey, you cited this person before', although I'm not like always trying to cite the same people, but when I did my knowledge tracing work in my first few years, [A] did come up a lot. So being able to see that is cool'' (P4). In addition, the \new{count} of papers was \new{helpful for} quantifying \new{known} connections: ``[A] authored four papers I cited. Wow, yeah, I knew that I cited him but I didn't have, like, any kind of quantification of it'' (P4).

\textit{B3-b. Scientist's Intellectual Impact.} A \new{specific} category of connections to research communities \new{is} impact, for example when a scientist's work is cited by others in the community to develop the research area further. Participants found citation-based messages useful for understanding their impact: ``When it says `cites one paper by you', I got excited. It also helped me contextualize this work better'' (P7); ``It (the recommended paper) cites me... this is something that I cannot immediately figure out (without this message) unless I opened up the paper and, like, control-F'ed my name'' (P6); ``The messages that talk about how it cites my papers get me really curious. And I'll almost always go look at it, to figure out how my work is being discussed and how my work influenced their work. And maybe think about future work.'' (P14).

\textit{B4. Mobilizing mental models of scientists' work and expertise areas.} Participants found author-based relevance messages especially useful when they had prior understanding of the featured authors' work. The mental model of scientist\new{s} was broadly represented as \new{a} combination of \new{their} research topics, frequently used epistemological approaches, and \new{their} seniority. Using mental models participants could better make sense of the recommended paper's contribution, its broader intellectual context, and inform their filtering decisions \new{as described in the following sections}.

\textit{B4-a. Readily available mental models of topical areas and understanding broader research context.} To many participants author names could be readily mapped as `specific topical areas' (P9) or `general research directions' (P14) which can be useful for filtering. Though less frequent than topical indexing, author names also signalled the quality of work: ``Papers from \new{[A]} are always pretty good, so I probably will read it.'' (P14). 
%P9 commented: ``a specific individual in your career that you've collaborated with will represent like a specific topical area... so seeing a relevance message featuring that individual will help map that specific topical area.'' 
%Similarly P14 said: ``if it's [Author], I know, it's probably about, like, model explanation, about human AI collaboration, could be about AI and crowdsourcing. That general direction... so it's both a social endorsement and also an explanation of the topic of the recommended item.'' Though less frequent than topical indexing, author names also signalled the quality of work: ``Papers from [Author] are always pretty good, so I probably will read it. It's also kind of related to my crowdsourcing work'' (P14). 
%Similarly P4 said ``Yeah, I guess if I see his name on a paper, I know, it's going to be about one of these two fields of knowledge tracing or intelligent tutoring systems. And it could be like, a new angle on or something, but it's like, you know, for sure one of the central ideas can be one of those two things.'' 
Mental models of authors extended beyond topical associations or quality signals and sometimes even helped participants understand \textit{a priori} in what concrete context the user needs present in recommended papers might arise. P6 described this phenomenon in the following quote:  ``I don't do ML work, but I can understand how there might be pain points in team communications around the ML model quality... so the \new{[A]}'s work I'm familiar with is in documentation and programmer support tools. So I can imagine how that goes for teams with multiple stakeholders, not all of them are technical, especially in places like [large technology corporation] that I believe \new{[A]} is at.'' % -- P6

% , P6 commented that seeing author names sometimes helped her understand the 
% add another quote about negative filtering that's NOT from P4

\textit{B4-b. Mental models for transfer.} \new{For indirect author-based messages,} participants could transfer their readily available mental models of the known indirect author to make an educated guess about the unfamiliar author. This was perceived as useful for mapping out `how ideas diffuse' (P2), `who's building off of the old but important work in the field' (P2, P6) or `working in an interesting intersection of fields' (P4, P9), \new{which is} currently `a nebulous and difficult task' (P2).
%P2 commented: ``To see how ideas diffuse, or how people follow up on certain ideas? Like who is citing who, and what is building off of what. But it's currently a nebulous and difficult task to do. I can see this message useful for that'' (P2).
%Similarly, P6 commented: ``seminal papers are obviously very important... but might not be the most recent. So maybe someone's done something based off of that seminal research that's more specific to the space that I'm interested in'' (P6). 
Furthermore the unfamiliar author\new{'s work} may be understood \new{through the mental model of the known author, for example as an indication of} `an important link that is missed' (P12). P4 commented: ``\new{[IA]} does some cognitive science stuff but he's not a huge cogsci guy... so maybe \new{[A]} does similar work to \new{[IA]} but like a combination of \new{[IA]}'s work with a more of a cogsci spin to it? Which is cool.'' (P4). Similarly P12 said: ``Okay this paper maybe is from a machine learning community I don't follow. But apparently [IA] cited a bunch of [A]'s papers, so maybe I'm missing something.'' (P12).

%  I don't know if [Author] and [Indirect Author] have a direct personal connection, but like [Author] I know has authored a lot of papers about diversity and open source and that's an area that [Author] is really interested in, so I can immediately understand what this connection is, just by looking at it, I can probably guess what six papers (that the Indirect Author= cited of the Author'' (P6) -- can we frame this as transfer

\textit{B5. Serendipitous author discovery.} \new{Author-based relevance messages also} led participants to pay attention to author names they otherwise may not have noticed, and serendipitously discover new and interesting authors `outside the radar' (P12). P10 said: ``A really interesting concept, but I don't recognize this author, let me check his publications... (clicks on the author link to check his profile) Okay wow, this is like a gold mine, I can easily spend an hour or two reading his papers.'' %(P10).
% Similarly P12 commented that ``(the messages lead to) like serendipity, I'm discovering something new that would otherwise have been completely outside of my radar.'' (P12)
Discovery of authors also happened in a more structured manner by transferring mental models from the known indirect author to the unfamiliar author. Participants found authors that they had not known before, but felt like they \textit{should} know, given the significant connection through the indirect author. Participants described that they could `picture where the indirect author is citing the unknown author' (P6), and that a certain number of citations from the indirect author represented `a good body of work by the unknown author' (P4) and a strong signal of how their research interests are aligned, which indicated the potential value of discovering the new author.
%P6 described ``Okay so yeah I can like picture probably where [Indirect Author]'s citing this person. So novice programmers... I do feel like I should know this person, though. I see API stuff, I see GitHub stuff, like oh no I really should know this person'' (P6). 
%P4 made a similar comment: ```16 different papers that [Indirect Author] has cited' makes me feel like I've probably cited them decently, and that person must do stuff that's obviously related to what [Indirect Author] does, which is somewhat related to what I do. So I probably should know this person (goes on to click on the link to the author page).'' -- P4.

\textit{B6. Judging potential value of a paper.} \new{Indirect author-based messages} were useful for vetting the value of the recommended papers. Participants described how knowing the scientist that they give credit to cited the author of the recommended paper multiple times gave them confidence that the recommended paper would also be useful to them. P14 described that this form of relevance implied a high chance of utility that he, too, could use the recommended paper to support claims in his papers, given how often his advisor \new{(who was the featured indirect author in the message)} cited the author. \new{For this reason, it mattered} whether the indirect author \new{was} someone whose work the scientist knows and gives credit to (``the fact that this mathematician that I give credit to is citing this paper, then that gives credit to the paper.'' -- P3), \new{otherwise the relevance may be} ignored. However, one participant commented that he would be interested in seeing a highly selective group of influential authors (``5--10 most cited researchers'' -- P8) in each subfield to be featured in the messages, regardless of personal connections. Sometimes indirect authors were useful for negative filtering, too \new{e.g., ``[A] is in like a CS education-y, a different field than me... I care about what she's doing, but it's a less cool factor for me'' (P4); ``[A] works with computationally very heavy mechanisms. Not what I'm interested'' (P10).} %Going to the opposite direction, sometimes indirect authors were useful for negative filtering due to the topical distance (P4, P10).
%P14 described how knowing the scientist that he gives credit to cited one of the authors of the recommended paper multiple times times gave him confidence that the recommended paper would be useful to him: ``actually think this is pretty interesting, because this means [Indirect author] used [Author]'s work to support his claims and in his papers. So that gives me more confidence about this paper being useful for my future work. Like, there's a high chance of me being able to use this to support my claims'' (P14). Similarly, P3 commented ``the fact that this mathematician that I give credit to is citing this paper, then that gives credit to the paper.'' P2 commented that the endorsement signal from indirect author messages were useful in conjunction with other signals such as the publication venue, while P5 described how it triumphs other sources of signal: ``The fact that there is this middleman which is a relevant person to me, is almost good enough for me to ignore the venues'' (P5). P8 wanted to see similar endorsement from influential authors in the field (``maybe you can use CSRankings.org to see who are the top authors in each subfield and feature connections to them?'' -- P8). Going in the opposite direction, participants also used author names as a negative filter: ``[Author] is in like a CS education-y, a different field than me. So I mean, I care about what she's doing, but... it's a less cool factor for me'' (P4); ``[Author] works in a computationally very heavy mechanism... not what I'm currently interested'' (P10).

These results uncover qualitative insights \new{that} complement \new{the effectiveness of messages in Study 1}. In addition, they also \new{show how} indirect author-based relevance messages \new{complemented the other two types of messages}. \new{Next,} we describe \new{the challenges involved with current messages.}

\xhdr{Types of Challenges} \textit{C1. Interpretability of indirect author-based messages.} \label{subsection:interpretation}
\new{Generating} relevance messages in a post-hoc manner could lead to incongruent information. While some participants regarded messages as something that `doesn't hurt to have' (P4) or `a weak signal' (P3) even when they were not found to contain particularly useful information, others \newn{indicated confusion when encountering a} relevance message for a paper that did not seem to relate much to their topical interest (``It gives me more confidence about the recommendation that there's a high chance that I can cite this paper too, but I don't know how it's related to me'' -- P14). Compared to citation- and direct author-based messages, which featured directly relevant information to users, indirect author-based messages were associated with `a steep learning curve' (P6) and subject to frequent misinterpretation and re-reading, though participants could get used to them after seeing a few messages \new{of the same type} (``The first time was confusing but getting the hang of it now'' -- P2). Th\new{is} perceived difficulty of interpretation was consistent with participants' subjective ratings on the \textit{Ease} question, in which indirect author-based messages scored significantly lower ($\mu=4.9, \bar{\sigma}=1.21$) than direct author-based messages ($\mu=5.9, \bar{\sigma}=1.04$) (two-tailed Welch's t-test, $t(22.74)=2.18, p=0.04$).

Part of the difficulty was anticipated by our research team given the nature of the message that involved second-degree relations between authors, and was proactively mitigated by \new{the second line text that described the} scientists' relation to indirect authors. This was perceived as helpful by 8 out of 14 participants (e.g., ``I can tell right away that it's my personal connection'' -- P1, Fig.~\ref{table:likert}, bottom), but only when they had some mental model of the indirect author (``I find the text marginally useful when I don't recognize the name'' -- P1). Participants also wished to see more contextual information that could remind them of `forgotten connections' (P2) or their own previous interaction data such as which paper they found interesting or had saved in the library, which led to the relevance messages being surfaced.

\textit{C2. Tension between more context and efficiency.} Unlike names found on a social network where users have real-world relations with each other, names extracted from the \textit{implicit} `social' network of authors may not be grounded in real-world relations and thus necessitate additional contextual information that helps users understand who the authors are and how they might be related.
Participants frequently mentioned wanting to see specific contextual information helpful for making sense of the relevance surfaced from the messages. P7 said: ``I think it could definitely use more detail. Because I don't know what I cited, from this context. I know she cited our paper. So now I am more interested, like, oh, what did she say?''. For P1, providing author names in the message alone was not nearly enough, given how he needed contextual information even for contacts with personal ties. P2 noted that while most names were unmemorable, their papers might be, and contextualizing and reminding users of their previous behaviors could help (``Is there a way to tell why I thought an author interesting before?'' -- P2). Other participants also wanted to see fine-grained citation context such as its salience (``it matters more if it was more of a key citation'' -- P11) or the section information.
%This led him to describe how additional topical associations might help: ``One problem is that I don't really remember authors too much. Even my contacts usually have notes describing what they do. So if there was an author name, if you could put a list of topics that that person is citing next to it... (that would be helpful)'' (P1). 
% Forgetfulness was  issue for P2: ``Is there a way to tell why I thought an author interesting before?'' (P2). 

Ultimately participants \new{thought} that additional context \new{could} fulfill their filtering needs, as P6 put it:     ``I know enough about \new{[IA]} to say that he has done some research I'm very interested in and then some that I'm less interested in... I guess he kind of has two camps of research, or two or more at least, and so I'm curious like, this person has authored six papers that \new{[IA]} cited... \textit{what camp did this \new{(paper)} fall into}, or like what papers is \new{[IA]} citing? Are they gender papers or otherwise, cuz that \textit{changes a lot how much I care.}'' %-- P6. 
When participants had a deeper understanding of the expertise areas of an author, they sometimes wanted to filter based on a subset of areas they cared about the most, as relations based on \newn{less central} areas of interest were less useful or irrelevant. Indirect authors may have done `a lot of great, but very diverse work' (P7) which may not be interesting to the user. 

Taken together, these comments suggest that there is a rich sub-space of citation and author information that can be used to further contextualize the relevance messages, and that it may be most effective when aligned with scientists' task-specific filtering needs. \new{However,} while additional context may help scientists reason about the relevance between authors and themselves, it may also increase the chance of information overload~\cite{wilson2008improving}, creating a tension between desiring more information and \new{also} wanting to quickly scan the recommendations in the email. Several participants, including P2, aptly described this potential tension: ``I usually want to scan as quickly as possible the titles... I feel like having to parse and switch back and forth between different kinds of relations to indirect authors will tire me.'' %-- P2.

\textit{C3. Evolving research topics and scientist's identities.} Another important \newn{identified} \new{issue} \newn{involves} how scientists' interests change over time, and that they may move away from old topics they themselves previously published in or move back to them with renewed interest later on. Depending on the changes, relevance extracted from citation graphs may become stale over time, and detect a `pioneering but universally cited' (P11) body of work as highly relevant, or relevance to `old research interests' \new{(3/14 participants)}. At the same time, participants also expressed the need to see recommendations related to their `past selves as scientists' (P14) for specific use cases like accepting review requests on those topics. Taken together, there \new{may be} high activation research areas at any \new{time} for scientists which constitute their current identities as scientists, \new{with several} `past identities' that consist of somewhat dormant, but occasionally reactivated topical areas. Accurately detecting such identities as scientists may be difficult or even impossible, yet an important aspect for systems aimed at surfacing current relevance to the users nonetheless.
%``high number of cites in messages may not always be helpful because this person for example was really, really significant in the field a long time ago, like the 90s. And when we cite papers we cite him again again.'' -- P11; ``I really love this paper. And the fact that they're citing my work. But it is not my current research.'' -- P7; ``I used to work in this topic but no longer with my current interest in the topic for my thesis.'' -- P8. However, participants were also interested in seeing recommendations related to their `past selves' because they sometimes had to wear a reviewer's hat in domains that they previously published in: ``This is relevant to my prior work, not my current interest but I still get a lot of review requests for it, so I still need to keep up to date with that literature'' -- P14. However, predicting how scientists conceptualize moving across different research interests over time and which of their publications fall into different `scientist's identities' can be a difficult task that challenges automatic extraction of citation-graph-based relevance. 
%  ``Yeah, definitely relevant, these two are relevant to my most cited papers in CSE Ed which I unfortunately don't do anymore.'' -- P4; 

\textit{C4. Trust.} Participants were more sensitive to potential errors in relevance between authors than in relevance based on citation. Because an important ingredient of effective relevance messages was participants' prior mental models of other scientists, for author-based messages this naturally invited conceptualization of relevance based on people and their intellectual legacy. For example, a high number of citations \newn{of an author} represented `a good body of work' for someone (P4) or a potential `advisor-advisee relationship' (P6), while a low number of citations could have meant `an up-and-coming' or a `junior' scientist in the field (P5). 

\new{T}his also meant higher sensitivity to how the relevance between authors was represented in messages, which could erode trust in the system when they suspected the relevance was not accurately quantified. P5 noted that using an exact quantification of how many papers authors have written or cited could be a `risky statement' because they were falsifiable and the margin of errors was small, especially for scientists that he knew well. Yet, systems that use citation graphs are subject to inevitable sources of errors, given the challenges from the scale and the speed of changes (e.g., \new{papers} may be published at \new{different} platforms such as ArXiv.org or conference proceedings at different \new{times}). One of the participants also noted a case where the relevance message was featuring a \new{very} different number of citations from what he expected, which he suspected was due in part to the author's deadname being incorrectly updated. As such, a challenge for systems that aim at inferring implicit social networks of authors from citation graphs is recognizing the increased sensitivity to the accuracy of author-based relevance and sufficiently fail-proofing or communicating the uncertainty of data associated with the representations of relevance to prevent erosion of trust.

\section{Implications for Design}
\begin{table*}[t]
    \centering
    \begin{tabular}{p{5cm} p{2cm} p{9.5cm}}
    \toprule
    \textbf{Design Implication} & \textbf{Areas} & \textbf{Summary} \\
    \midrule
    Goal-Centric Interpretability & C1, C2, B[2--6] & Messages need to be tailored to scientists' specific goals in order to work as a useful source of signal.\\
    \addlinespace[.1cm]
    Task-Centric Configurations & C[2--4] & Messages need to support task-specific configuration needs: \textit{here and now} vs. \textit{there and then}. \\ %Tuning alert frequency is already good. scientists may be interested in and need different contextual information personalized for their tasks. Skimming -- want to do this configuration once and forage multiple times when emails are received 
    \addlinespace[.1cm]
    Dynamic Scientist Identities & C3, B4, B5 & Author-based relevance needs to better capture and organize relevance through multiple scientist identities and temporally changing communities. \\
    \bottomrule
    \end{tabular}
    \caption{\textbf{Overview of Design Implications.} Using the benefits and challenges identified in Section~\ref{subsection:qualitative_results}, we synthesized design implications for future relevance messaging approaches that combine citation graphs and implicit social networks of authors.}
    \label{table:design_implication}
    \vspace{-2em}
\end{table*}
We further propose three main \new{design areas} for future relevance messages that combine citation graphs and implicit social networks of authors to support scientists' sensemaking, filtering, and discovery (\new{overview in} Table~\ref{table:design_implication}).

\xhdr{Goal-Centric Interpretability} Our participants used relevance messages for a host of diverse yet interrelated benefits. They saw authors featured in the messages as topical representations of their mental models of other scientists, and used them to contextualize the recommended paper and the connections between authors. Messages also helped the participants discover new and interesting authors that they otherwise might not have paid enough attention to notice. These specific use cases suggest that future strategies of relevance messaging need to be tailored to different goals scientists have in mind. In addition, scientists frequently engaged in multiple use cases while reviewing a single email suggesting that support for fluid switching between different goals while interacting with the content in a single email, or throughout multiple emails over a period of time may be important. 

One issue with the template-based uniform messaging approach adopted here is that although \newn{it} may \new{benefit} efficient scanning, th\new{is} efficiency comes largely from the uniformity of the \new{phrases} used in the messages. This results in generic, rather than specific descriptions of relations. Participants in our study also alluded to this, by suggesting alternative message designs focused on specific aspects of their needs such as ``convey the strength of topical similarity'' (\new{3/14 participants}), ``directly say a new author is worth checking out'' (\new{5/14 participants}), and ``describe the topical context of the author-author connections'' (\new{9/14 participants}). \new{Furthermore,} in exploratory scenarios under time pressure where multiple sources of information compete for user's attention and judgement of relevance, the amount of effort required for goal-specific interpretation of messages may be prohibitively high. Addressing this in future designs could greatly improve the effectiveness of relevance messages by reducing the gap of interpretability.

% In addition, generic messages require scientists to first decide which kind of benefits they want to get out of them, and could further confuse them on how they are related to their goals. This additional step due to the lack of messaging specificity may ironically increase the {ir}relevance of the messages and make them ignored, or leave them actively interfere with specific user goals by drawing users' (ultimately) wasted attention into them. For exploratory scenarios under time pressure where multiple sources of information compete against each other for user's attention and judgement of relevance, the amount of effort required for goal-specific interpretation of messages may be prohibitively high. Addressing this in future designs could greatly improve the effectiveness of relevance messages by reducing the gap of interpretability.

One potential design space is to suggest possible goals to scientists, and adaptively changing the message templates based on their choice. Because this additional discovery \new{step} of available options may \new{add cost to the} messaging strategies, recommender systems may proactively present variations of the messages and provide scientists a selection mechanism with minimal cognitive load. In addition to \new{message presentation and selection}, alternative measures of relevance may be computed and used to quantify the strength of relevance, \new{supporting scientists' choice of the messaging strategy.} For example, the strength of relevance may be computed as a normalized count of connections the recommended paper has to a user-curated feed, and further augmented with topical similarity \new{to each paper on the feed} by leveraging readily available NLP techniques.

In sum, \new{an important} challenge here for designers is to provide tailored support for a diverse set of goals that newly emerged with relevance messages. Making an assumption about default user goals \new{and supporting only static messages corresponding to them, or generic messages that pose an interpretability gap} may ultimately lead to user frustration and abandonment.

% many aspects of the recommended papers and their authors, it is important to ...

% explaining the Why -- challenge of post-hoc generation and interpretability. Also motivated by the diverse kinds of benefits of messages. -- do we want to support them all? do we want to focus on a specific kind or two? do we want to support user ability to 

% Challenges

% Design implication 1: Provide a way to dig deeper into the meaning of the messages, provide additional supporting messages to guard against misinterpretation
% Concrete solution ideas: A question mark, when hover’ed over that shows the explanation of message 
\xhdr{Task-Centric Configurations} 
% filtering
% discovery
Scientists need better support for their specific task context and ways to effectively manage their limited attention. The fundamental challenge is that scientists frequently experience information overload, and while personalized neural recommenders can help they can still result in an overwhelming amount of information to comb through. The three high-level tasks that scientists in our study commonly performed were filtering, sensemaking with mental models of other scientists, and discovery. Scientists expressed needs for additional support in configuring the messages to filter based on blacklisted/whitelisted authors, in prioritizing messages with specific citation context and importance, and in dialing up the salience of important relations between authors featured in the messages.

While participants appreciated existing modalities of longer-term configurations such as tuning the frequency of alert emails (\new{mentioned by 5/14 participants}) and the ability to receive new paper recommendations based on a set of curated papers or certain authors, relevance messages \new{also} introduced several user needs for additional configurations that apply within an individual email. For example, participants wanted to configure messages to feature additional `academic status' (\new{3/14 participants}) and `topic cards' (\new{4/14 participants}) next to unfamiliar author names to quickly get a sense of their work, and similarly `citation context' (\new{8/14 participants}) next to citation-based messages. \new{We saw that sometimes recommendations in an email acted as a launchpad for separate  discovery loops, for example by clicking author links to jump to the author details pages of people they wanted to learn more about (8/14 participants), though there were differences among the participants in terms of when they intended to actively engage in the discovery loops (e.g., as soon as they open the details page (3/14 participants) vs. storing them as open tabs for review at a later time (5/14 participants)).} The challenge for designers here is how to notice when filtering-oriented tasks evolve into discovery and vice versa, such that task-specific configurations can be effectively and adaptively supported. The changing nature of the tasks noted here also nods to the findings in information seeking behaviors (e.g.,~\cite{russell1993cost,white2009exploratory}) that involve alternating between broad foraging and focused exploiting phases. From our observations, participants may need two distinctive types of configuration support for tasks that are `here and now,' which are characteristic to scientists' \textit{in situ} filtering and micro-discovery needs within an alert email, vs. `there and then,' which are more relevant to configurations expected to last longer thus re-configured less frequently.

\xhdr{Dynamic Scientist Identities}
Relevance messages that leverage implicit social networks of authors should better reflect scientists' evolving research interests. One potential design space is in bootstrapping temporal shifts in topical interests by identifying sub-groups of scientists from prior publications and interaction data (e.g., curated research feeds and libraries around specific topics). Inferring temporally changing community structures from citation-graphs and implicit social networks of authors may also be useful for better supporting user needs in understanding how scientists are connected in a community and how within- and cross-community impact occurs. Another potential design space is in supporting multiple \textit{scientist identities} for individual authors. While publication history can be a useful source of signal, it also consists of a collection of related research areas that the scientist has previously embarked on, some of which may no longer be relevant to other scientists' interests. Bootstrapping the scientist-topic structures by segmenting time periods with high topical consistency from the publication history may be helpful. In addition, increasing the recency and topical relevance by simple discounting of older publications may be effective. 

Support for multiple scientist identities can also help mitigate the message--recommendation incongruence. Featuring messages on paper recommendations that are topically distant can confuse users into thinking that the recommended paper is highly relevant. Yet, due to post-hoc generation, messages that feature relevance through a\new{n overall} thread of research may augment papers that belong to \new{its} topically distant \new{segments}. In some cases participants were left wondering how to make sense of the conflicting signals of `a high confidence of utility signal from the message' (P14) but low perceived topical relevance of the recommendation.
% Scientists wanted additional ways to support their mental models of authors in relevance messages. 
%One of the Participants from the lab study described that the need for additional contextual information  when they noticed incongruence between recommendations and their \textit{current} research interests, and when relevance messages were featured on those recommendations deemed as topically distant. The issue seemed to occur due to a simple assumption that our relevance model was based upon: scientists' publication and interaction data remain temporally and topically consistent. This assumption held true for some as shown in the use cases of our study, but also resulted in misalignment for others when interests shifted significantly over time. Ignoring dynamic changes in temporal and topical dimensions and the resulting mismatches with user expectations can lead to reduced relevance and decreased user adoption of relevance messages.

\new{Therefore}, designers who wish to use author-based relevance information to augment scientific recommendations need to consider dynamic changes happening within an individual scientist's career trajectory as well as the community-level shifts over time. Capturing and organizing author-based relevance using such structures has the potential to better orient scientists in the multifaceted and dynamic space of relevance.

\section{Discussion}

\xhdr{\new{Inferring targeted social networks for messages}} \new{Previous work on social explanations has shown that social explanations can be persuasive, but they might be only a secondary effect to the people's inherent preferences and quality expectations about the recommended items~\cite{sharma2013social}. This was consistent with our observations --- the engagement benefit of messages for scientists was also influenced by other context such as topical alignment, freshness, and quality cues from the content and metadata of the paper (title, author names, publication venues, etc.).}

\new{A prominent difference between the social messages developed in this work and those in prior work lies in how targeted the network leveraged for message generation was for the recommendation task. On one end \newn{of the spectrum}, explicit social networks leveraged in~\cite{sharma2011network,sharma2013social} (e.g., Facebook) may capture real-world friendships among the users, but may also be less targeted for the task of recommending music, as friendship relations between two people encompass more than just similar tastes in music. On the other end, our inferred network of authors may lack the real-world relationships among the users, but may still capture task-specific relations about what different scientists read and how they build on each other's work. Our study results confirmed that the relations featured in messages indeed were relevant in this regard, and also useful because they activated existing mental models of the familiar authors featured in messages to improve their understanding of the recommended papers or the novel authors. There may also be a hybrid of networks for generating social messages. For example,~\cite{guy2009personalized} explored one approach to aggregating two types of relations: those from an explicit, public social network and others from a more targeted, within-organization intranet, in order to find most relevant and trustworthy co-workers to support the given recommendation task. The use of different types of social networks raises interesting open questions as to how they differ in terms of message persuasiveness (i.e., how much explanations boost user engagement) and informativeness (i.e., user satisfaction from consuming the recommended item). \newn{Answering these questions} requires studies that directly measure user perceptions before and after consuming the recommendations augmented with various types of messages pulled from different kinds of networked relations, that go beyond the persuasiveness focus (operationalized by click-through rates) of this work. Understanding these questions may bring new insights as to how different types of social network relations may be combined or used to facet one another to improve the persuasiveness and informativeness of recommendations.}

\xhdr{\new{Revisiting the contrast between informativeness and persuasiveness}} \new{Prior work has also contrasted the informativeness and persuasiveness of recommendations~\cite{bilgic2005explaining}, and investigated how applying social explanations often did not simultaneously increase both~\cite{sharma2013social}. This contrast is interesting to revisit in light of our findings for a few reasons. First, the structure of trust may be more targeted in the inferred scholarly network than in explicit social networks. For example,~\cite{sharma2013social} found that users trusted good friends more and perceived social explanations featuring them as more persuasive than when featuring random friends, but this persuasiveness was not highly correlated with their ultimate liking of the music after consumption. However, this finding may be due to the specific tie strength used to categorize friends in the work (e.g., good vs. random) that was less targeted to the music recommendation task. In contrast, using an inferred scholarly network of authors from publications and their citation network may have a benefit of surfacing more targeted relations useful for judging the ultimate utility of the recommendations, even though the relations may not be grounded in the real-world social relations. Future work may \newn{test} this hypothesis, and also further explore our findings around the challenges related to interpreting and trusting the inferred relations. }

\new{Second, the ultimate informativeness measure may be more nuanced in the domain of scholarly recommendations due to potential differences in their utility curve. Unlike other domains such as music or movie recommendations, scientists are both consumers and producers of the recommendations, rather than consumers alone. Therefore, the users are incentivized by and actively look for recommendations with both immediate (e.g., a reference to cite in the manuscript currently writing) and longer term utility goals (e.g., papers in new domains for future research) in mind. This suggests depending on the task context, each user may exhibit a different informativeness utility curve of recommendations and more or less openness to diverse recommendations that have different kinds of informativeness. Lastly, the scientific community consists of both lateral (e.g., peers) and vertical relations (e.g., direct advisory relations, or distant advisory relations through academic lineage) that users may leverage to interpret messages. Therefore, the notion of trust here is not limited to the immediate closeness among the friends as leveraged in previous work, and supports the use of intermediate author relations (e.g., `I want to know what my peers or my advisor's former collaborators are working on', `I trust B's recommendations because he's a well-known expert in the field'). Our findings from indirect author-based messages mediated by middle authors opened a new design space and also uncovered challenges for identifying trusted and preferred middle authors \newn{from whom} users might appreciate recommendations. Important questions remain open for future investigation, such as how the systematic choice of middle authors shifts the system-wide visibility of work (e.g., `is the effect similar to crowning an author with a prestigious, status-conferring prize?'~\cite{reschke2018status}), whether it concentrates, reinforces, or creates new pathways to the power of persuasion by specific authors (e.g., `how does the dynamic differ or align with the megaphone effect~\cite{mcquarrie2013megaphone}?'), and if so, how \newn{the goals of} trust and fairness may be balanced in the selection algorithm.}

%intuitively this make sense because. .. how the the limited ability of social explanations in helping users accurately predict their actual liking of the content recommended to} them~\cite{sharma2013social}, c

\xhdr{Limitations and broader impacts} \new{The methods developed here were evaluated on a particular search engine and within the email alert context. However, given \newn{that our approaches are agnostic} to the underlying recommender algorithms, they may be applied to other settings, for example, to supplement recommendations found on social media such as Twitter. Future work investigating whether our findings generalize to such settings may be important.} More investigation is also needed to understand the full impact of interventions like ours to ensure fairness with respect to other important attributes, including \new{author} gender \new{and ethnicity, and the full scope of the impact of the increased engagement, including} unintended consequences such as coming at the expense of other important activities. In addition, whether our messages influence diversity of consumption (e.g.,~\cite{anderson2020algorithmic}) remains an important open question. 
 %Furthermore, while our relevance messages may be seen as a form of persuasion, (i.e., via social proof or authority principles \cite{cialdini2009influence}), we were very careful to avoid any forms of deception or purposeful misrepresentation, by focusing on providing only the best-available factual information. 

Our operationalization of author reputation is also limited in several ways. While we looked at multiple ways of aggregating h-index across authors of a paper (max and average), other aspects of authorship such as the reputation of the lead author or the last author may be important to consider. We also considered only the changes in papers clicked directly from the email, whereas our intervention could potentially also influence downstream papers clicked in the future.

Finally, the real-world nature of \new{Study 1} introduce\new{d} some unavoidable measurement error. For example, due to the wide range of devices \new{and event context}, it was not possible to ensure that we accurately detected all events such as email opens.
It is also very challenging to automatically extract knowledge graph elements such as citations and disambiguate authors, particularly for newly published papers which may not be accompanied by publisher metadata, and errors in this process introduced additional noise into the deployment.
\new{While it is unlikely that significant biases were introduced into experimental conditions in a systematic way because participation was randomized, improving the data coverage and accuracy from the deployment setting will be valuable for surfacing the full scopes of effects.}

\section{Conclusion}
Scientists today are faced with a daunting yet fundamental task of staying on top of the large, rapidly growing literature. With little support from existing tools to know why the recommended papers might be worth their attention to read, scientists are forced to wrangle long, monotonous lists of recommendations---and perhaps quit in the process. To better support scientists' broad information needs and mitigate the issue of scarce relevance signals, we designed and empirically tested two kinds of graph-based relevance messages by finding connections from who they know to what they read.
%through who the scientists know and what they read for personalized paper recommendations. 
Our large-scale, real-world online deployment study revealed empirical evidence that relevance messages are an effective means for engaging scientists. To further increase their benefits, we designed and implemented a third kind of messages via \new{an inferred scholarly network and featuring relations mediated by middle authors that the user may trust}. From a controlled lab study with 14 scientists we gained qualitative insights into the utility of our relevance messages as well as their challenges. Finally, we synthesized a set of \new{future implications for design}, \newn{which} aim to use \new{inferred social network} relevance to engage scientists. We envision a future in which scientists are delighted to keep up with academic literature through personalized paper recommendations that help them attend to authors they know, discover new \new{interesting} authors found from what they have read or interacted with in the past, and provide many additional engaging and helpful signals \new{that feed into a positive loop of further improving the recommender systems.}

%%
%% The acknowledgments section is defined using the "acks" environment
%% (and NOT an unnumbered section). This ensures the proper
%% identification of the section in the article metadata, and the
%% consistent spelling of the heading.

\begin{acks}
% Anonymized for submission
This project is supported in part by NSF Grant OIA-2033558, NSF RAPID award 2040196, ONR grant N00014-21-1-2707, and the Allen Institute for Artificial Intelligence (AI2). The authors thank Alex Schokking, Alex Buraczynski, Paul Sayre, Cecile Nguyen, Sebastian Kohlmeier, and Rodney Kinney for their advice on and help with engineering and Kyle Lo for his advice on the analysis of experimental results. We also thank the anonymous reviewers for their constructive feedback. Finally, this work would not have been possible without our study participants. %\bug

\end{acks}

%%
%% The next two lines define the bibliography style to be used, and
%% the bibliography file.
\bibliographystyle{ACM-Reference-Format}
\bibliography{main}

\subsection*{Appendix A. Post-hoc analyses on the potential sources of biases of randomization} \label{appendix:posthoc-analysis-randomization}
\begin{figure}[h!]
    \vspace{-1em}
    \begin{minipage}[t]{.225\textwidth}
        \centering
        \includegraphics[height=0.23\textheight]{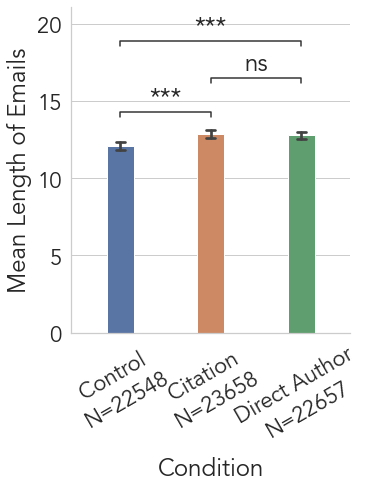}
        \vspace{-1em}
        \caption{The avg. length of emails were slightly shorter in Control ($\mu=12.1, \bar{\sigma}=20.30$) than the other two conditions ($\mu=12.9, \bar{\sigma}=21.74$ in Citation and $\mu=12.8, \bar{\sigma}=17.36$ in Direct Author), and the difference was less than one paper.}
        \label{fig:mean_email_length}
    \end{minipage}
    \quad
    \begin{minipage}[t]{.225\textwidth}
        \centering
        \includegraphics[height=0.23\textheight]{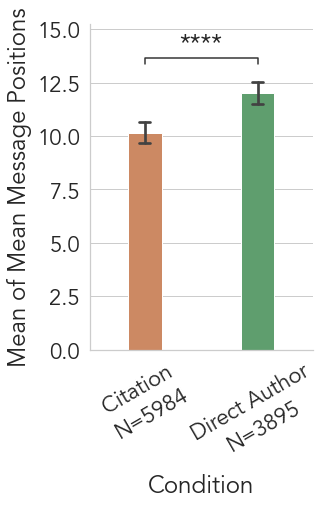}
        \vspace{-1em}
        \caption{The mean of mean rank positions of messages in emails was significantly higher in Direct Author than in Citation, \new{suggesting receivers of direct author messages had to read more through the list in the email to see the messages.}}
        \label{fig:mean_of_mean_message_position}
    \end{minipage}
    \caption{Examining potential biases of randomization.}
    \vspace{-1em}
\end{figure}

% \begin{figure}[h!]
%     \subfloat[The average length of emails were slightly shorter in Control ($\mu=12.1, \bar{\sigma}=20.30$) than the other two conditions ($\mu=12.9, \bar{\sigma}=21.74$ in Citation and $\mu=12.8, \bar{\sigma}=17.36$ in Direct Author), and the difference was less than one paper.]{%
%     \includegraphics[height=0.238\textheight]{figures/mean_num_papers_per_email.png}
%     \label{fig:mean_email_length}}
%     \quad
%     \vspace{-1em}
%     \subfloat[The mean of mean rank positions of messages in emails was significantly higher in Direct Author than in Citation, \new{suggesting receivers of direct author messages had to read more through the list in the email to see the messages.}]{%
%     \includegraphics[height=0.238\textheight]{figures/mean_of_mean_ranks.png}
%     \label{fig:mean_of_mean_message_position}}
%     \caption{Examining potential biases of randomization.}
%     \vspace{-1em}
% \end{figure}

We further ran post-hoc analyses on two potential sources of biases -- the average length of emails and the average position of relevance messages -- to ensure the validity of random assignment. Systemic differences in these biases between conditions may significantly change user engagement, and have potential to invalidate the interpretation of the results on engagement. For example, longer emails may fatigue users and cause them to drop off more easily. At the same time, longer emails may also provide users more paper recommendations to look at, and as a result an increased click engagement. On the other hand, the average position of relevance messages would positively impact user discovery and attention, because messages placed higher (towards the top of the email) are more easily discovered.

These potential sources of biases appear unlikely to have changed the directions of the observed effects on engagement. Though the length of emails \textit{did} significantly differ between conditions and on average the emails in the Citation and Direct Author conditions were longer than in Control, this difference was less than a paper in both cases ($\Delta_{\text{Citation} - \text{Control}}=0.78$ and $\Delta_{\text{Direct Author} - \text{Control}}=0.69$).
The average number of paper recommendations clicked per email was $\mu=0.10, \bar{\sigma}=0.680$; the expected number of clicked emails from the difference is then $0.78\times0.10=0.078$ for the Citation condition and $0.69\times0.10=0.069$ for the Direct Author condition. However, the average number of clicked emails in both treatment conditions was still higher after accounting for the expected surpluses of clicked papers, $\mu=0.24, \bar{\sigma}=0.936$ (Citation) and $\mu=0.36, \bar{\sigma}=1.262$ (Direct Author). Between the Citation and Direct Author conditions, the average position of relevance messages indexed from the top of the email was smaller (i.e. closer to the top of the email) in the Citation condition ($\mu=10.1, \bar{\sigma}=20.55$) than in the Direct Author condition ($\mu=12.0, \bar{\sigma}=17.72$). This suggests that if the discoverability of relevance messages were controlled for, Direct Author messages may lead to an even greater increase in user engagement.

\subsection*{\new{Appendix B. Difference-in-Differences (DiD) analysis on the email open rates}}
\new{We further examined the effect of messaging strategies on future engagement, while accounting for the effect of familiarity and habit-forming with email alerts over time. Though Fig.~\ref{fig:open_rates} showed the overall differences in email open rates between conditions, it is possible that there was a significant time effect on users' decision to open an incoming alert email, given their previous experiences with them, such as habitual opening. This would likely create a positive trend-line for the baseline email open rates over time, thus making the estimation of true effects of our messaging strategies harder to obtain. We applied Difference in differences analysis to account for such effect. \emph{DiD} is a technique often used in econometrics and social sciences to derive causal inferences from observational, panel data~\cite{meyer1995natural}. We used efficient linear probability estimation method based on linear regression models. This has an additional benefit of producing directly interpretable coefficients~\cite{ai2003interaction}.}

\new{We applied two symmetric linear regression models, each corresponding to one messaging strategy (i.e. Citation or Direct Author). The regression model's dependent variable was the binary open rates (1: Email was opened, 0: Otherwise), and the predictive variables included whether the email was sent within the first two weeks of the experiment (Early Exposure), whether the email was in the treatment condition (vs. Control), and the interaction of the two. The number of emails in the Early Exposure vs. Late Exposure groups was roughly equal, with the biggest difference occurring in the Citation condition ({11,459} emails in the early exposure group vs. {12,199} in the late exposure group).}

\begin{figure}
    \centering
    \includegraphics[width=.3\textwidth]{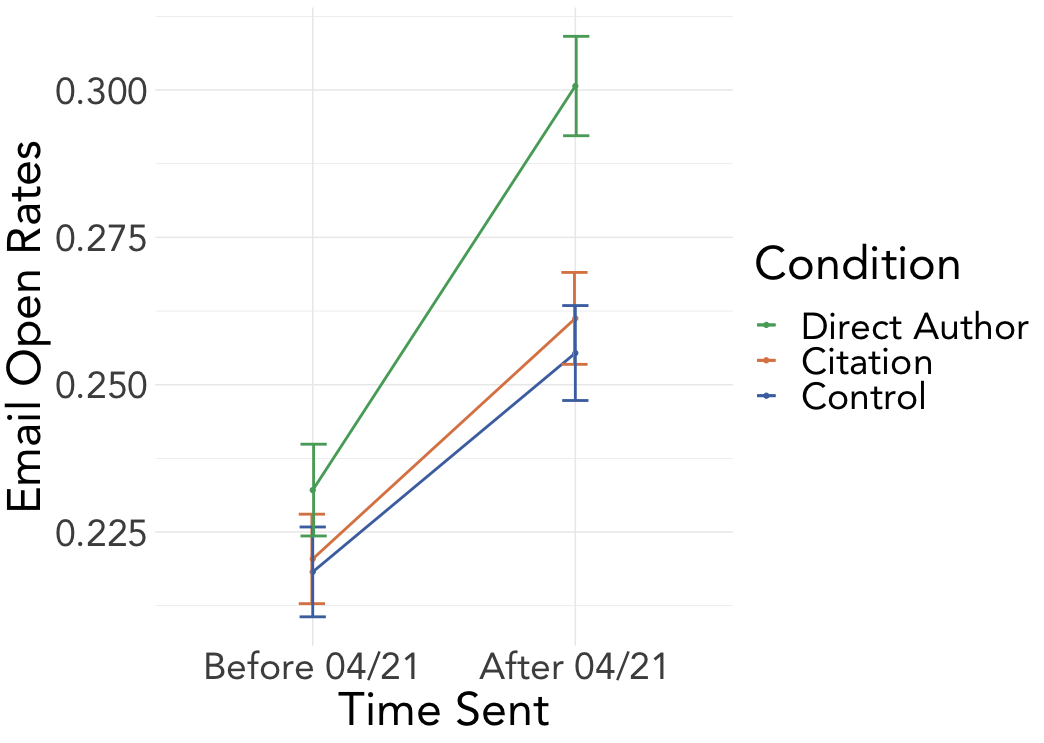}
    \vspace{-1em}
    \caption{\new{Mean email open rates by condition shows a significantly steeper slope in Direct Author over Control, suggesting its effectiveness in boosting the open rate after accounting for the baseline increase over time.}}
    \label{fig:did-plot}
\end{figure}

\new{The result of our analyses showed that indeed there was a significant time effect on the open rates in both conditions. Specifically, Early Exposure was significantly \textit{negatively} associated with the open rates, suggesting that users became increasingly `habitual' with opening new alert emails over time (Table~\ref{table:did-citation},~\ref{table:did-direct-author}). After accounting for this time effect, the effect of messaging was significant only in the Direct Author condition ($p<0.001$), but not in the Citation condition ($p=0.64$), further lending support to the effectiveness of the author-based messaging strategy (Fig.~\ref{fig:did-plot}). Specifically, later Direct Author messages induced a 30\% increase in email open rates relative to the first two-week exposure period.}
\begin{table}
\begin{tabular}{@{\extracolsep{.5pt}}c c c c}
    \toprule
     & Coef. & SE & $p$ \\
    \midrule
    (Intercept) & 0.256 & 0.004 & *** \\
    Early Exposure & -0.037 & 0.006 & *** \\
    Message (\textit{Citation}) & {0.006} & {0.006} & {0.29} \\
    Interaction & -0.004 & 0.008 & 0.64 \\
    \bottomrule
    \end{tabular}
    \captionof{table}{\new{\textbf{The result of linear regression with Citation messages and Control.} Citation messages did not increase email open rates after accounting for the baseline increase over time ($p=0.64$). ***: $p<0.001$, **: $p<0.01$.}}
    \label{table:did-citation}
    \vspace{-2.5em}
\end{table}
\begin{table}
    \begin{tabular}{@{\extracolsep{.5pt}}c c c c}
    \toprule
     & Coef. & SE & $p$ \\
    \midrule
    (Intercept) & 0.256 & 0.004 & *** \\
    Early Exposure & -0.037 & 0.006 & *** \\
    Message (\textit{Direct Author}) & {0.045} & {0.006} & {***} \\
    Interaction & -0.031 & 0.008 & *** \\
    \bottomrule
    \end{tabular}
    \captionof{table}{\new{\textbf{The result of linear regression with Direct Author messages and Control.} Direct Author messages significantly increased email open rates in addition to the baseline increase over time ($p<0.001$). ***: $p<0.001$, **: $p<0.01$.}}
    \label{table:did-direct-author}
    \vspace{-2.5em}
\end{table}

\subsection*{Appendix C. Empirical relationship between the predictive variables and click-through rates \new{(\textsc{ctr})}}
\begin{figure*}[h!]
    \vspace{-1em}
    \begin{minipage}[t]{.3\linewidth}
        \centering
        \includegraphics[height=0.15\textheight]{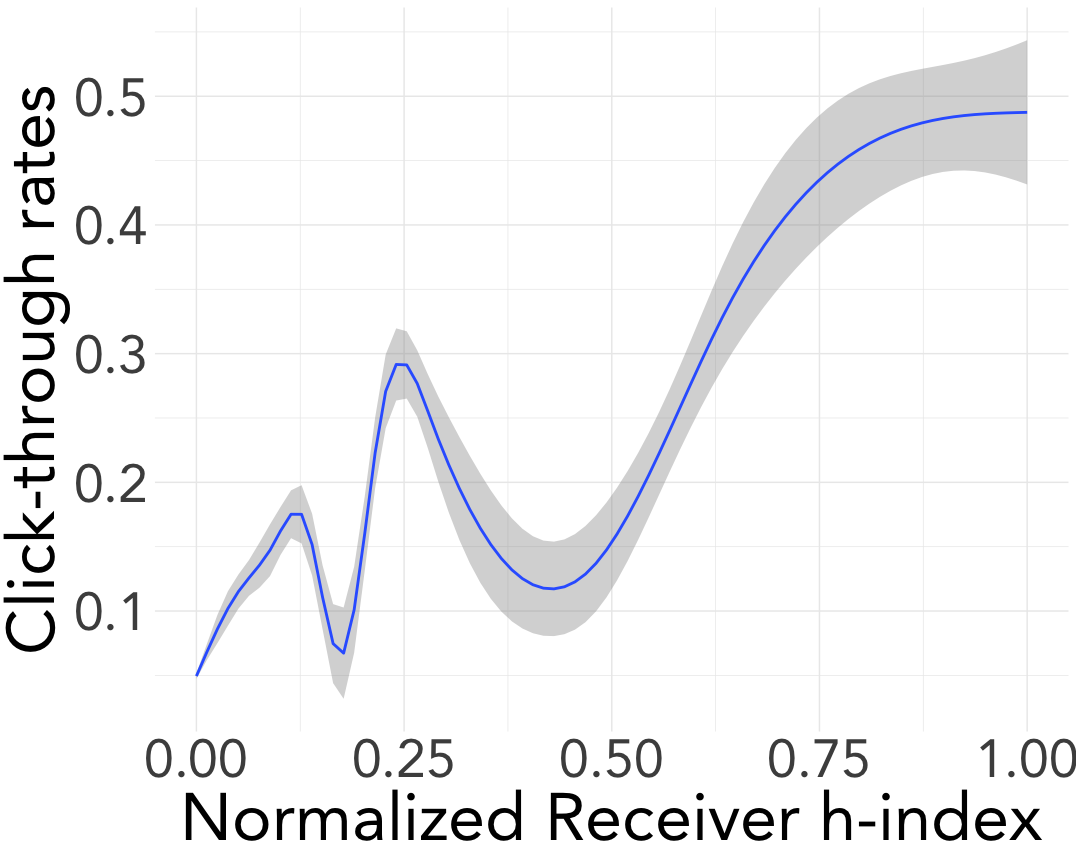}
        \vspace{-1em}
        \caption{\new{The LOWESS plot suggested an overall positive correlation between receiver h-index and \textsc{ctr}.}}
        \label{fig:clicked_by_h}
    \end{minipage}
    \quad
    \begin{minipage}[t]{.6\linewidth}
        \centering
        \includegraphics[height=0.15\textheight]{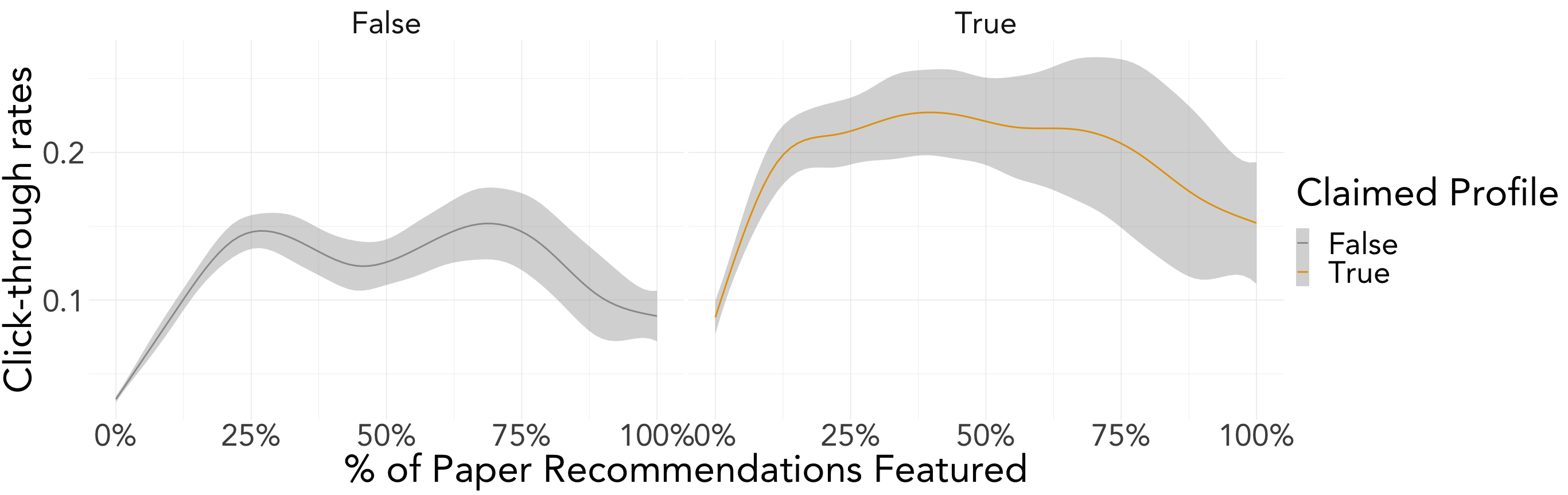}
        \vspace{-1em}
        \caption{\new{Faceted LOWESS plots suggested inverted U-shaped curve relationships between \% of paper recommendation featured and \textsc{ctr}, with an overall higher level of \textsc{ctr} for claimed profile users.}}
        \label{fig:clicked_by_claimed_profile}
    \end{minipage}
    \caption{LOWESS plots of \textsc{ctr} data faceted by different covariates.}
    \vspace{-1em}
\end{figure*}

% \begin{figure*}[t!]
%     \subfloat[\new{The LOWESS plot suggested an overall positive correlation between receiver h-index and \textsc{ctr}.}]{%
%     \includegraphics[height=0.15\textheight]{figures/email_clicked_normalized_h_index_combined.png}
%     \label{fig:clicked_by_h}}
%     \quad
%     \subfloat[\new{Faceted LOWESS plots suggested inverted U-shaped curve relationships between \% of paper recommendation featured and \textsc{ctr}, with an overall higher level of \textsc{ctr} for claimed profile users.}]{%
%     \includegraphics[height=0.15\textheight]{figures/email_clicked_claimed_profile_facet_combined.png}
%     \label{fig:clicked_by_claimed_profile}}
%     \vspace{-1em}
%     \caption{LOWESS plots of \textsc{ctr} data faceted by different covariates.}
% \end{figure*}
\begin{figure}[h]
    \centering
    \includegraphics[width=.22\textwidth]{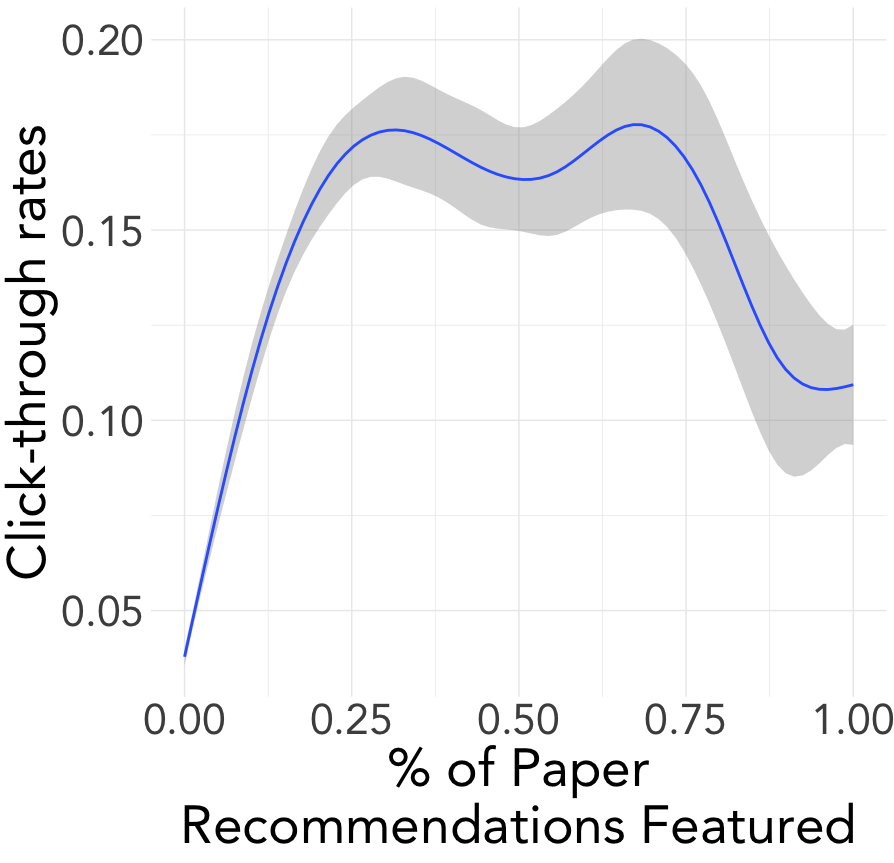}
    \vspace{-1em}
    \caption{\new{The LOWESS plot of \textsc{ctr} data by the \% of paper recommendations in the email featured with relevance messages suggested an approximately inverted U-shaped relationship.}}
    \label{fig:clicked_by_fraction_featured}
\end{figure}
The Locally Weighted Scatterplot Smoothing (LOWESS) plot showed a roughly inverted U-shaped curve between \textsc{ctr} the \% of paper recommendations featured with \textit{any} type of relevance message, with a greater fraction of treated papers resulting in dramatically more engagement, up to a point (between 25-50\% treated) above which engagement falls off (Fig.~\ref{fig:clicked_by_fraction_featured}). Furthermore, \new{it also showed \textsc{ctr} was roughly positively correlated with the receiver's h-index (fig.~\ref{fig:clicked_by_h}), and whether they claimed a profile (fig.~\ref{fig:clicked_by_claimed_profile}), showing users with claimed profiles showing overall higher levels of \textsc{ctr}.}

\subsection*{Appendix D. Variance Inflation Factor analysis to determine factors that should be pruned due to significant collinearity with others} \label{appendix:GLMM}
We performed Variance Inflation Factor (VIF) analysis~\cite{allison1999multiple} on the predictor variables (\textit{Claimed Profile}, \textit{\% Featured}, and \textit{\# of Total Papers}) because they are assumed to have parallel causal influences on \textit{Email Clicked}. The VIF value for each predictor variable is typically obtained by first regressing it against all others in the set, then computing the $1/\left(1-R^2\right)$. For example a VIF of 1.5 tells us that the variance of the predictor variable is 50\% greater than would be the case if no collinearity was present. Following~\cite{johnston2018confounding}, we used 2.5 as a threshold for the existence of significant collinearity among the predictor variables. The result showed that the highest VIF from the predictor variables was 1.14 (\textit{Claimed Profile}) hence we proceeded without pruning any variable from the model.

\subsection*{Appendix E. Likelihood Ratio Test for modeling the relationship between \new{\textsc{ctr}} and \% of recommendations featured with messages}
We hypothesized that an optimal \% of papers within an email featured with author-based relevance messages (\textit{\% Featured}) for click-through is neither too few -- which may lead to under-utilization, nor too many -- which may lead to overwhelming the users. Investing this curvilinear relation requires inclusion of a quadratic variable in the model. The descriptive pattern from the data also suggested this hypothesis (Fig.~\ref{fig:clicked_by_fraction_featured}). Using locally weighted scatterplot smoothing (\textsc{LOWESS}), we observed that the peak click-through happened somewhere between 25\% and 50\% of the papers featured with author-based relevance messages, and quickly decreased outside this range. Along with the descriptive pattern, we also performed a pairwise Likelihood Ratio Test (LRT) on two models, one with only the linear \% Featured term and the other with only the quadratic $\left(\text{\% Featured}\right)^2$ term, in order to further test the soundness of introducing the quadratic term. The result showed that the reduction of deviance from introducing the quadratic term is more than what we would have expected to see if the beta coefficients for them were 0 ($\chi^2(1)=19.9, p=8.1\times10^{-6}$), thus we proceeded with introducing the quadratic term and building Model 1:\footnote{We did not perform an additional check on the VIF when including the quadratic term, despite its potentially significant collinearity with the linear term. See~\cite{o2017dropping} for more discussion on the relevant topic.}.
\begin{align*}
    g(E[y]) &= \beta_0 + \gamma_j + \beta_1x_1 + \beta_2x_2 + \beta_3x_3 \\
    g(E[y]) &= \beta_0 + \gamma_j + \beta_4{x_1}^2 + \beta_5x_2 + \beta_6x_3
\end{align*}
where $y$=\new{\textsc{ctr}}, $\beta_1$ representing the fixed effects from the \textit{\% Featured} ($x_1$), $\beta_2$ representing the fixed effects from the \textit{Claimed Profile} variable ($x_2$), $\beta_3$ representing the fixed effects from the \textit{\# of Total Papers} ($x_3$), $\beta_4$ representing the fixed effects from the quadratic (\textit{\% Featured})$^2$, and similarly for $\beta_5$ and $\beta_6$. Random intercepts $\gamma_j \sim \mathcal{N}(0, \sigma_{\gamma}^2)$ were introduced for participants $j$'s. We used the logit link $g(p)=\text{log}(p/(1-p))$ to model the engagement as a Bernoulli variable.
\begin{table}
    \centering
    \begin{tabular}{@{\extracolsep{4pt}}r c c c}
    \toprule
    & Coef. & SE & $p$ \\
    \midrule
    (Intercept) & -8.20 & 0.278 & *** \\
    $\text{\% Featured}$ & 3.17 & 0.600 & *** \\
    $\left(\text{\% Featured}\right)^2$ & -3.00 & 0.680 & *** \\
    \# of Total Papers & 0.01 & 0.038 & 0.81 \\
    \bottomrule
    \end{tabular}
    \caption{\new{Regression analysis using Model 1 predicted significant curvilinear changes on \textsc{ctr} from the \% of papers featured with author-based relevance messages while the total number of recommendations in the email was not a significant predictor. *** indicates significance at $p<0.001$.}}
    \label{table:glmm1}
    \vspace{-2em}
\end{table}

\begin{figure*}[t!]
    \begin{minipage}[t]{0.24\linewidth}
        \begin{subfigure}[t]{\linewidth}
            \includegraphics[height=0.18\textheight]{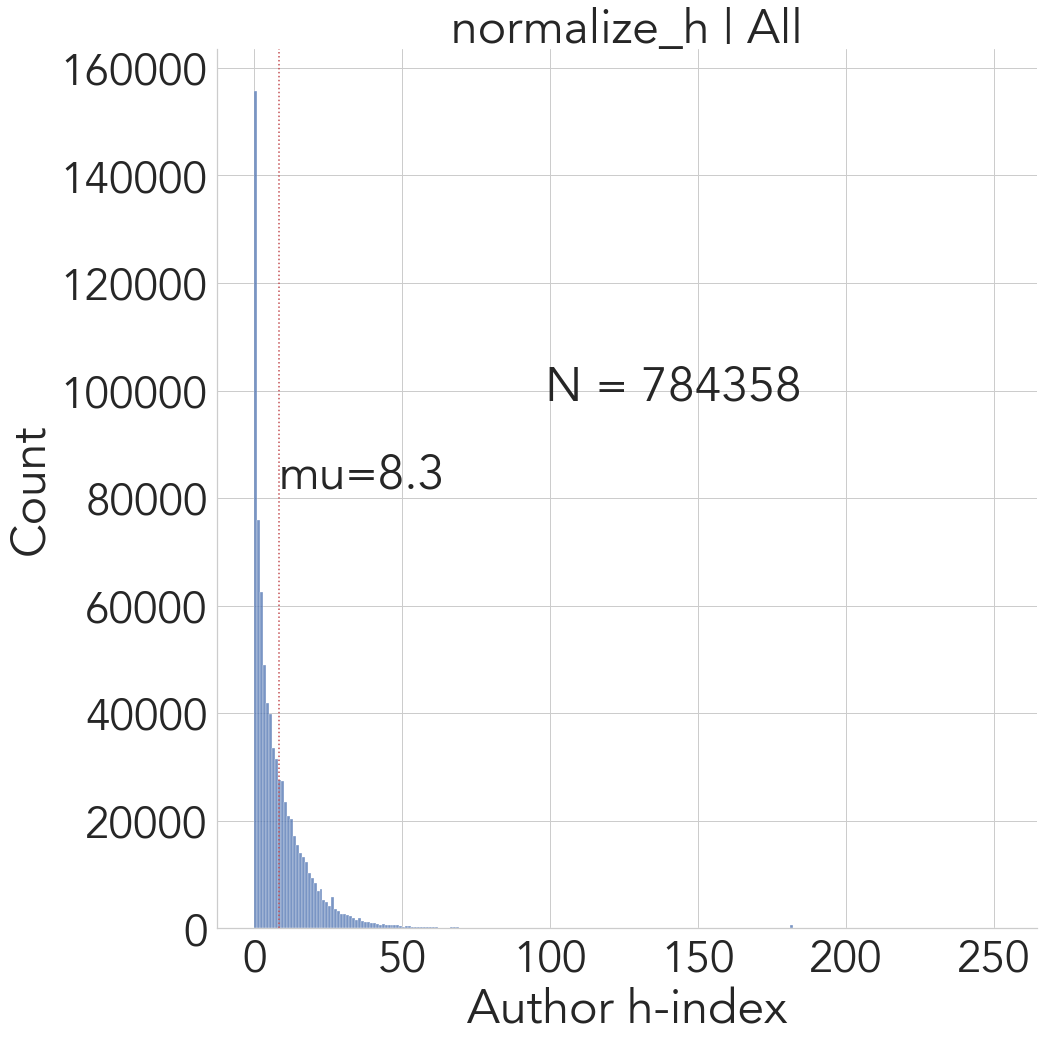}
            \caption{The `background' distribution of normalized author h-indices of all paper recommendations sent out in all alert emails in all conditions. $\mu=8.3, \bar{\sigma}=11.14$ (median=$5.0$).}
        \end{subfigure}
    \end{minipage}
    \quad
    \begin{minipage}[t]{0.72\linewidth}
        \begin{subfigure}[t]{\linewidth}
            \includegraphics[height=0.18\textheight]{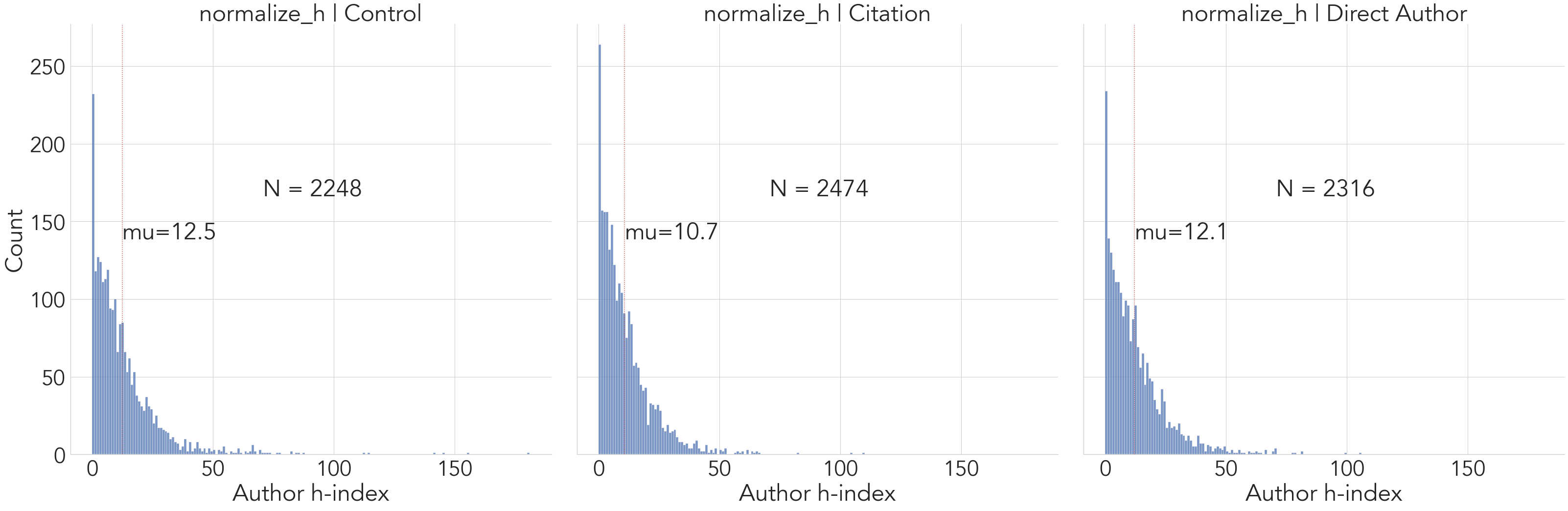}
            \caption{The distribution of normalized author h-indices of clicked papers in each condition. The average increased to $\mu=12.5, \bar{\sigma}=14.25$ (Control, median=$8.7$); $\mu=10.7, \bar{\sigma}=10.87$ (Citation, median=$8.0$); and $\mu=12.1, \bar{\sigma}=12.11$ (Direct Author, median=$9.0$). \new{The increase of the average h-index over the background distribution was significant, suggesting that users considered high status authors as a signal for deciding whether to click on a paper by default. At the same time, citation messages reduced the h-index relative to control, suggesting its effect of guiding user attention to lesser known authors. Direct author messages and control did not differ significantly.}}
        \end{subfigure}
    \end{minipage}
    \vspace{-1em}
    \caption{Analysis showed users on average clicked on papers with authors with higher average h-index, and featuring relevance messages did not shift user attention to only the papers with high-profile authors.}
    \label{fig:h_distribution_normalized}
    \vspace{-1em}
\end{figure*}

% \begin{figure*}[t!]
%     \subfloat[The `background' distribution of normalized author h-indices of all paper recommendations sent out in all alert emails in all conditions. $\mu=8.3, \bar{\sigma}=11.14$ (median=$5.0$).]{%
%         \includegraphics[height=0.18\textheight]{figures/all_normalized_h.png}
%         \label{fig:all_normalized_h}}
%     \quad
%     \subfloat[The distribution of normalized author h-indices of clicked papers in each condition. The average increased to $\mu=12.5, \bar{\sigma}=14.25$ (Control, median=$8.7$); $\mu=10.7, \bar{\sigma}=10.87$ (Citation, median=$8.0$); and $\mu=12.1, \bar{\sigma}=12.11$ (Direct Author, median=$9.0$). \new{The increase of the average h-index over the background distribution was significant, suggesting that users considered high status authors as a signal for deciding whether to click on a paper by default. At the same time, citation messages reduced the h-index relative to control, suggesting its effect of guiding user attention to lesser known authors. Direct author messages and control did not differ significantly.}]{%
%         \includegraphics[height=0.18\textheight]{figures/treatment_normalized_h.png}
%         \label{fig:clicked_normalized_h}}
%     \vspace{-1em}
%     \caption{Analysis showed users on average clicked on papers with authors with higher average h-index, and featuring relevance messages did not shift user attention to only the papers with high-profile authors.}
% \end{figure*}

The result of Model 1 is shown in Table~\ref{table:glmm1}. This regression tells us whether the probability of \new{\textsc{ctr}} changes significantly if we increased the \% of papers featured with author-based relevance messages, or changed the number of paper recommendations included in each email. Consistent with our hypothesis, we found a significant curvilinear relationship between the \% of papers featured in the email and user engagement. The coefficients of both the linear \% Featured and its quadratic terms were significant ($p<0.001$), and the signs were in the opposite direction, with a negative beta coefficient for the quadratic term. This validates the empirically observed inverted U-shaped curve relationship on \new{\textsc{ctr}}. We also found that the normalized length of email was not a significant predictor of user engagement, \new{which suggests} the conflicting effects of longer emails: overwhelming users (negative) vs. providing more paper links to click (positive). We extend Model 1 with additional predictive variables and account for them in our regression analysis using Model 2 (Section~\ref{subsubsection:study1-frequency}).

\subsection*{Appendix F. Repeat analysis on the fairness of visibility (Section~\ref{section:h-index-analysis}) using normalized author h-index} \label{appendix:max-h-analysis}
The repeat analyses of author h-index distributional shifts using normalized author h-index per paper recommendation (\new{Fig.~\ref{fig:h_distribution_normalized}; see also} Section.~\ref{section:h-index-analysis}).

\end{document}